\documentclass[12pt]{article}   	% use "amsart" instead of "article" for AMSLaTex format
\usepackage{geometry}                		% See geometry.pdf to learn the layout options. There are lots.
\usepackage{graphicx}				% Use pdf, png, jpg, or eps§ with pdflatex; use eps in DVI mode
\usepackage{amssymb}
\usepackage{enumitem}
\usepackage{amsthm}
\usepackage{setspace}

\usepackage[authoryear, sort]{natbib}
\usepackage{xcolor}

\usepackage{booktabs}
\usepackage{array}
\usepackage{multirow}

\usepackage{float}
\floatstyle{plaintop}
\restylefloat{table}

\usepackage[format=plain, labelfont=bf, singlelinecheck=off]{caption} 

\usepackage{subcaption}
\usepackage{amsmath}
\usepackage{verbatim}
\numberwithin{equation}{section}

\usepackage{etoolbox}
\makeatletter
\patchcmd{\@begintheorem}{\itshape}{}{}{}
\patchcmd{\@opargbegintheorem}{\itshape}{}{}{}
\makeatother

\usepackage{algorithm}% http://ctan.org/pkg/algorithms
\usepackage{algpseudocode}% http://ctan.org/pkg/algorithmicx
\newlength{\continueindent}
\usepackage{etoolbox}

\usepackage{xr} %% for cross-reference 
\algnewcommand{\algorithmicgoto}{\textbf{go to}}%
\algnewcommand{\Goto}[1]{\algorithmicgoto~\ref{#1}}%

\newtheorem{theorem}{\textsc{{Theorem}}}[section]

\newcommand{\bfb}{\mathbf{b}}

\newcommand{\bfh}{\mathbf{h}}

\newcommand{\bfs}{\mathbf{s}}
\newcommand{\bfp}{\mathbf{p}}

\newcommand{\bfr}{\mathbf{r}}

\newcommand{\bfx}{\mathbf{x}}
\newcommand{\bfy}{\mathbf{y}}
\newcommand{\bfz}{\mathbf{z}}

\newcommand{\bfA}{\mathbf{A}}

\newcommand{\bfC}{\mathbf{C}}
\newcommand{\bfD}{\mathbf{D}}

\newcommand{\bfF}{\mathbf{F}}
\newcommand{\bfG}{\mathbf{G}}
\newcommand{\bfH}{\mathbf{H}}
\newcommand{\bfI}{\mathbf{I}}

\newcommand{\bfK}{\mathbf{K}}

\newcommand{\bfP}{\mathbf{P}}

\newcommand{\bfR}{\mathbf{R}}

\newcommand{\bfT}{\mathbf{T}}

\newcommand{\bfW}{\mathbf{W}}
\newcommand{\bfX}{\mathbf{X}}

\newcommand{\bfZ}{\mathbf{Z}}

\def\mNNGP{\textrm{NNGP}}

\newcommand{\bfell}{\boldsymbol \ell}
\newcommand{\Ell}{\boldsymbol \ell}

\newcommand{\bfbeta}{\boldsymbol \beta}
\newcommand{\bfeps}{\boldsymbol \epsilon}

\newcommand{\bftheta}{\boldsymbol \theta}
\newcommand{\bfmu}{\boldsymbol \mu}

\newcommand{\bfsigma}{\boldsymbol \sigma}
\newcommand{\bftau}{\boldsymbol \tau}

\newcommand{\bfSigma}{\boldsymbol \Sigma}
\newcommand{\bfLambda}{\boldsymbol \Lambda}
\newcommand{\bfOmega}{\boldsymbol \Omega}

\newcommand{\calL}{\mathcal{L}}

\expandafter\def\expandafter\normalsize\expandafter{%
    \normalsize
    \setlength\abovedisplayskip{6pt}
    \setlength\belowdisplayskip{6pt}
    \setlength\abovedisplayshortskip{6pt}
    \setlength\belowdisplayshortskip{6pt}
}

\allowdisplaybreaks

\usepackage{hyperref}

\usepackage{geometry}
\usepackage{setspace}
\usepackage{authblk}

\usepackage{titlesec}
\titleformat{\subsection}
  {\bfseries\slshape}
  {\thesubsection}
  {1em}
  {}
 
 \titleformat{\subsubsection}
  {\slshape}
  {\thesubsubsection}
  {1em}
  {}

\newcommand{\blind}{1}

\begin{document}
\singlespacing

\if1\blind
{
\title{\bf Computer Model Emulation with High-Dimensional Functional Output in Large-Scale Observing System Uncertainty Experiments}

\author[1,2]{Pulong Ma\thanks{\emph{Correspondence to: Pulong Ma, Postdoctoral Fellow at the Statistical and Applied Mathematical Sciences Institute and Duke University, 79 T.W. Alexander Drive, P.O. Box 110207, Research Triangle Park, NC 27709. Email: pulong.ma@duke.edu.}}}
\author[3]{Anirban Mondal}
\author[4]{Bledar A. Konomi}
\author[5]{Jonathan Hobbs}
\author[6]{Joon Jin Song}
\author[4]{Emily L. Kang}
\affil[1]{Statistical and Applied Mathematical Sciences Institute}
\affil[2]{Duke University}
\affil[3]{Case Western Reserve University}
\affil[4]{University of Cincinnati}
\affil[5]{Jet Propulsion Laboratory, California Institute of Technology}
\affil[6]{Baylor University}
\date{}							% Activate to display a given date or no date
  \maketitle

} \fi

\if0\blind
{

  \bigskip
  \bigskip
  \bigskip
  \begin{center}
    {\LARGE\bf Computer Model Emulation with High-Dimensional Functional Output in Large-Scale Observing System Uncertainty Experiments}
\end{center}
  \medskip
  
} \fi

\begin{singlespace}
\vspace{-1.0cm}
\begin{abstract} 
Observing system uncertainty experiments (OSUEs) have been recently proposed as a cost-effective way to perform probabilistic assessment of retrievals for NASA's Orbiting Carbon Observatory-2 (OCO-2) mission. One important component in the OCO-2 retrieval algorithm is a full-physics forward model that describes the mathematical relationship between atmospheric variables such as carbon dioxide and radiances measured by the remote sensing instrument. This forward model is complicated and computationally expensive but large-scale OSUEs require evaluation of this model numerous times, which makes it infeasible for comprehensive experiments. To tackle this issue, we develop a statistical emulator to facilitate large-scale OSUEs in the OCO-2 mission with independent emulation. Within each distinct spectral band, the emulator represents radiances output at irregular wavelengths via a linear combination of basis functions and random coefficients. These random coefficients are then modeled with nearest-neighbor Gaussian processes with built-in input dimension reduction via active subspace. The proposed emulator reduces dimensionality in both input space and output space, so that fast computation is achieved within a fully Bayesian inference framework. Validation experiments demonstrate that this emulator outperforms other competing statistical methods and a reduced order model that approximates the full-physics forward model.   
\end{abstract}

\end{singlespace}
%\vfill

%\newpage 
%\doublespacing
\onehalfspacing
%\singlespacing

\setstretch{1.15} %doublespacing

\section{Introduction} \label{sec: intro}

%% background of the problem in the context of UQ
With space-based observations, remote sensing technology provides a wealth of information for understanding geophysical processes with unprecedented spatial and temporal coverage. Quantitative inference for the global carbon cycle has been bolstered by greenhouse gas observing satellites. NASA's Orbiting Carbon Observatory-2 (OCO-2) collects tens of thousands of observations of reflected sunlight daily. These observed spectra, or radiances, are used to infer the atmospheric carbon dioxide (CO$_2$) at fine spatial and temporal resolution with substantial coverage across the globe \citep[e.g.,][]{boesch2017}. Estimates of atmospheric CO$_2$ are computed from the observed radiances using an inverse method known as a retrieval algorithm. The mathematical and computational framework for the retrieval is problem-specific but often involves an optimization with a physical forward model of moderate computational complexity. OCO-2 has adopted a particular retrieval methodology known as ``optimal estimation (OE)'' \citep{Rodgers2000} in remote sensing science. The resulting estimates of geophysical quantities of interest are called retrievals. 

\begin{figure}[H]
\makebox[\textwidth][c]{\includegraphics[width=1.0\textwidth, height=0.15\textheight]{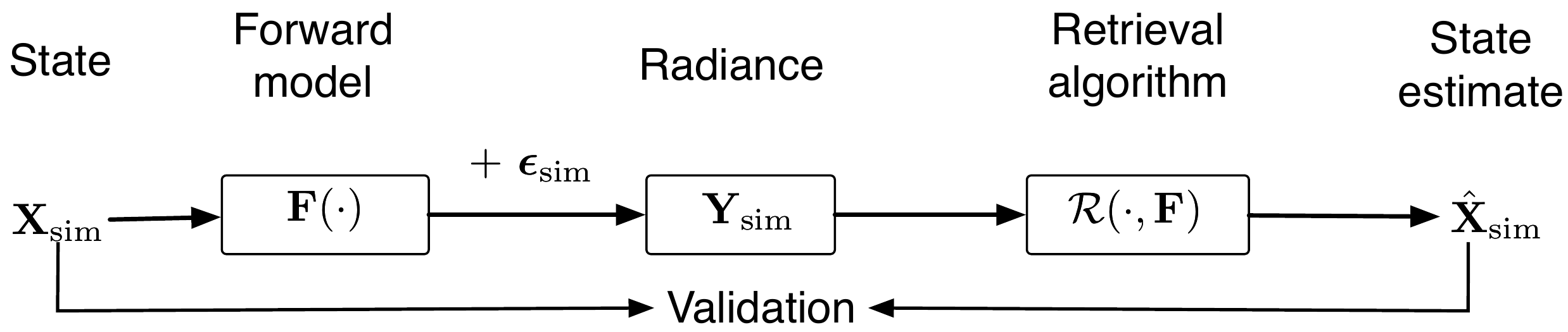}}
\caption{Diagram of a basic observing system uncertainty experiment adapted from \cite{Turmon2019}. The state is drawn from a probability distribution that represents the real world. The forward model describes the mathematical relationship between the state and noiseless measurements. $\bfeps_{\text{sim}}$ represents the instrument measurement errors in the OCO-2 satellite. Given radiance measurements, the retrieval algorithm generates estimates of the state, which are called retrievals. As long as the synthetic states and the transformations implemented in OSUEs are accurate depictions of the real-world processing chain, the statistical properties of the retrieved estimates $\hat{\bfX}_{\text{sim}}$, relative to the synthetic truth, $\bfX_{\text{sim}}$, will mirror those of the actual retrievals relative to the true state.}
\label{fig: OSUE}
\end{figure}

Probabilistic assessment and uncertainty quantification of remote sensing retrievals are crucial to the success of using these remote sensing data to reveal valid scientific findings and to answer scientific hypotheses appropriately. However, different from many other disciplines, it is infeasible to perform physical experiments to study the quality of remote sensing retrievals thoroughly because a representative ground truth of atmospheric variables is usually lacking; if there were such a ground truth, there would not have been the need to launch satellites to obtain information for these atmospheric variables from space. Therefore, \cite{Turmon2019} suggest using simulation-based experiments, known as \emph{observing system uncertainty experiments} (OSUEs), to quantify various sources of uncertainty in probabilistic assessment of remote sensing retrievals. Such experiments are supposed to be cost-effective, because OSUEs are based on simulation and can be executed for a range of observing conditions for which ground truth may not be available. Figure~\ref{fig: OSUE} shows a basic diagram to compare the retrieval to the synthetic true atmospheric state in an OSUE for remote sensing retrievals. \cite{HobbsBraverman2017}  present an OSUE and use it to study the impact of uncertain inputs on the distribution of the retrieval error in OCO-2. 

In order to thoroughly study uncertainty in the retrievals and the impact of parameters in the retrieval algorithm, it is necessary to perform large-scale OSUEs in which we need to perform simulation and evaluate the forward model $\bfF(\cdot)$ (Figure~\ref{fig: OSUE}) many times over a large spatial domain at various specifications of the model inputs. However, this forward model $\bfF(\cdot)$, designed to represent the complicated physical relationship between atmospheric variables and radiances, usually involves radiative transfer models and thus is computationally expensive, making the computational costs of large-scale OSUEs prohibitive. \cite{HobbsBraverman2017} present a surrogate model that is computationally efficient and substitute this new model for the full-physics forward model in order to make large-scale OSUEs practical. This surrogate model in \cite{HobbsBraverman2017}, referred to as a \emph{reduced order model} (ROM) hereafter, preserves some key physical laws in the original $\bfF(\cdot)$, referred to as the \textit{full-physics} (FP) forward model, but uses {a low-dimensional representation of the atmospheric state and} simplified numerical routines to solve the nonlinear equation of radiative transfer. Readers are referred to \cite{HobbsBraverman2017} for details of this ROM and the retrieval algorithm for OCO-2.

In this work, we focus on developing an efficient statistical emulator for the FP forward model $\bfF(\cdot)$ to facilitate probabilistic retrieval assessment via large-scale OSUEs. There is a well-established paradigm and vast literature on constructing statistical emulators for many expensive computer models \citep[e.g.,][]{Bayarri2007, Conti2009, Higdon2008, Mak2018, Sacks1989, Santner2018, Guillas2018, Welch1992}, but developing a statistical emulator for the OCO-2 FP forward model presents a few challenges: (1) The outputs of the FP forward model are radiances at irregular wavelengths, varying across spatial locations. Moreover, these radiance outputs from the FP forward model are at up to a total of 1016 wavelengths for the three near infrared spectral bands, termed the O$_2$-A band, the weak CO$_2$ band and the strong CO$_2$ band, respectively, but there are often values missing at some wavelengths irregularly due to the instrument's characteristics and its interaction with its environment in low-Earth orbit. Besides, these three spectral bands have completely non-overlapping wavelengths. (2) The input of the FP forward model in OCO-2 is a 66-dimensional vector. Such high-dimensional input can pose problems in emulator construction, especially for Gaussian process (GP) emulation, and thus dimension reduction is necessary \citep[e.g., ][]{Constantine2014, Constantine2014MC, Constantine2016, Liu2017}. (3) We have about $n = 10,849$ runs from the FP forward model based on which we need to build a statistical emulator. This size of data can cause computational issues for a GP emulator, since parameter estimation and prediction will require storing and factorizing matrices of size $10,849\times 10,849$. Note that in this article we use $10,849$ simulation runs to demonstrate our methodology but in general millions of FP forward model runs are available from previous simulations, which can be incorporated in the study. 

This article provides a unified framework to overcome these three challenges in order to construct a computationally efficient statistical emulator for large-scale OSUEs in the OCO-2 mission. In particular, motivated by \cite{Bayarri2007} and \cite{Higdon2008}, we treat the radiances as functional outputs and perform functional principal component analysis (FPCA) \citep{ramsaysilverman2005} to achieve dimension reduction. Similar to the principal component analysis (PCA) approach in \cite{Higdon2008}, FPCA is used as a data-driven approach to obtain a basis representation of the output; the associated weights are then modelled using Gaussian processes. Unlike PCA, which is mainly designed for outputs at a regularly-spaced grid without missing values, FPCA can handle outputs over irregularly spaced and varying wavelengths with missing values via basis functions of functional forms. \cite{Bayarri2007} use fixed wavelet basis functions placed over a dyadic grid for functional representation of the output. In our approach, B-spline basis functions are used for functional representation of the output, then the principal components are obtained as functions of wavelength using the data-driven FPCA. To reduce the dimensionality of input space, we choose to use the active subspace method \citep{Constantine2014, Constantine2017} because the gradient of the FP forward model in OCO-2 is available, and the active subspace method tends to give comparable or better results compared to other input dimension reduction methods including sufficient dimension reduction \citep{Cook1994, Cook2009} and effective dimension reduction \citep{Li1991} as demonstrated in literature \cite[e.g.,][]{Constantine2014, Constantine2017, Liu2017}. 

To alleviate the computational difficulty related to large training samples in GP emulation, we adopt the nearest neighbor Gaussian process \citep[NNGP;][]{Datta2016} developed in {the area of} spatial statistics and extend it to the context of computer model emulation. Specifically, we formulated two statistical emulators that allow a nonseparable covariance structure between input space and radiance output within each distinct spectral band and simultaneously allow separable covariance structures across different spectral bands. Here the nonseparable covariance structure between input space and radiance output is achieved through basis-function representation within each distinct spectral band, while the separable covariance structure between input space and radiance output across different spectral bands is achieved through two approaches: the one with independent emulation for each spectral band, and the one with a separable covariance model across different bands. The first emulator is referred to as a \emph{band-independent} emulator or simply \emph{independent} emulator, in which the FPC scores are modeled with independent NNGPs with distinct model parameters both within each distinct spectral band and across different spectral bands. As the physics of molecular absorption of CO$_2$ can allow strong correlation for radiances in weak CO$_2$ and strong CO$_2$ bands, we also develop a \emph{band-separable} emulator or simply \emph{separable} emulator, in which independent NNGPs are assumed for FPC scores within each distinct spectral band, but a separable covariance structure is assumed between input space and FPC weights in weak CO$_2$ and strong CO$_2$ bands to account for the cross-covariance between these two bands. Our proposed separable emulator is an upgraded version of the classic separable model in \cite{Conti2010} with the capability of handling large-scale computer experiments. For parameter estimation, we also propose to integrate out latent random variables and location-scale parameters. The resulting posteriors from these two emulators can be sampled using standard Markov chain Monte Carlo (MCMC) algorithm efficiently with fast convergence and mixing, since the posterior is only a function of correlation parameters and nugget parameter. This inference approach can avoid the slow convergence issues in the sequential updating MCMC using latent spatial random variables proposed by \cite{Datta2016}. We further demonstrate that the resulting emulators based on the NNGP can achieve efficient inference within a fully Bayesian framework and give satisfactory performance compared to the emulator based on the local Gaussian process approximation \citep{Gramacy2015}  and the ROM in \cite{HobbsBraverman2017}.

This article is organized as follows. Section~\ref{sec: motivating data} describes the overall picture of OSUEs for probabilistic assessment of remote sensing retrievals with the focus on the OCO-2 mission and introduces the OCO-2 data that are used to build the proposed emulators. Section~\ref{sec: Model} introduces the statistical framework of building emulators for the FP forward model in OCO-2 mission. In Section~\ref{sec: results}, we compare the proposed emulator with other statistical methods and validate the statistical emulators with a reduced order model. Section~\ref{sec: conclusion} is concluded with discussion and possible extensions.

\section{Probabilistic Assessment of Retrievals} \label{sec: motivating data}
This section gives a description of the probabilistic assessment of retrievals via observing system uncertainty experiments and the OCO-2 data. 

\subsection{Observing System Uncertainty Experiments}
In NASA's OCO-2 mission, the OCO-2 satellite collects the radiances data that are used to produce geophysical quantities of interest (QOI) such as the total column mole-fraction of CO$_2$ (known as XCO2 in the remote sensing community) via a retrieval algorithm, where the XCO2 is a weighted average of the estimated state vector $\hat{\bfx}$ with weights given by a location-specific pressure weighting function \citep{boesch2017}. XCO2 can be further used to estimate sources and sinks of CO$_2$ via flux inversion models \citep[e.g.,][]{Patra2017, Gurney2002}. Thus, the uncertainty propagated to these downstream science investigations is inherited from the estimated state vector $\hat{\bfx}$ \citep{cressiejasa}. Uncertainty quantification (UQ) has been a major focus of the OCO-2 mission and carbon cycle science community. 

In the remote sensing community, including OCO-2, the statistical formulation of the retrieval inverse problem often invokes a Bayesian formulation for retrieving atmospheric QOIs from noisy measurements of radiance spectra from satellite instruments. The formulation assumes a multivariate Gaussian prior distribution on the state $\bfx$ and a data model of the form 
\begin{align} \label{eqn: forward model}
\bfy & = \bfF(\bfx, \bfb) + \boldsymbol{\epsilon}, %\\
%\boldsymbol{\epsilon} & \sim \mathcal{N} (\mathbf{0}, \boldsymbol{\Sigma}_{\mathbf{\epsilon}} ), 
\end{align}
where $\bfx$ is the state vector assumed with a prior distribution.  $\bfb$ is a vector of the corresponding fixed model parameters that is needed in order to run the FP forward model $\bfF(\cdot)$. $\bfeps$ is the instrument measurement error that is often assumed to be a Gaussian distribution with zero mean and covariance matrix $\bfSigma_{\epsilon}$. The retrieval algorithm used in the OCO-2 mission aims at finding posterior mode of $\bfx$ given the radiance spectra $\bfy$, denoted by $\hat{\bfx}$, using the Levenberg-Marquardt optimization algorithm. 
Other than these quantities, the retrieval algorithm also depends on the convergence criterion and number of iterations used in the optimization algorithm. For detailed introduction and implementation of the operational retrieval algorithm in OCO-2; see \cite{boesch2017}. 

To quantify the uncertainty associated with the estimated state vector $\hat{\bfx}$, the remote sensing community often treats the forward model $\bfF(\cdot)$ as fixed and sets $\mathbf{b}$ according to knowledge of the underlying physics. While it is fixed in the retrieval, $\mathbf{b}$ is not perfectly known and is a source of uncertainty in the inverse problem (i.e., inverse UQ part) \cite{ConnorEtAl2016}. The standard assessment of retrievals on $\hat{\bfx}$ is to validate $\hat{\bfx}$ against ground truth such as ground-based CO$_2$ estimates collected by the Total Carbon Column Observing Network (TCCON) \citep{Wunch2011}. The TCCON network covers a variety of locations and geophysical conditions, but its coverage does have limitations. Further, the TCCON CO$_2$ estimates are subject to uncertainty, although they are generally found to have higher precision than OCO-2 retrievals \citep{zhangtechnom}.  

The OE retrieval framework allows for an analytical estimate of the posterior variance based on a linear approximation \citep{Rodgers2000}. This estimate is produced frequently in the remote sensing community for many observing systems including OCO-2, but it only accounts for the uncertainty in the radiance vector $\bfy$ and ignores the impact of misspecification of the forward model as demonstrated by \cite{HobbsBraverman2017}. In contrast,  \cite{Turmon2019} propose to use OSUEs in Figure~\ref{fig: OSUE} for uncertainty quantification of retrievals. The OSUE has the capability to account for uncertainties arising from the information flow from a synthetic state vector to the estimated state vector. Such experiments are cost-effective and can be used to quantify uncertainties at various stages in controlled experiments. This framework is detailed by \cite{Braverman2019} to guide practical implementation for uncertainty quantification of physical state estimates derived from remote sensing observing systems. It provides an \emph{end-to-end} uncertainty quantification for retrievals. 

From an uncertainty quantification perspective, the standard uncertainty estimates in the remote sensing community only quantify the uncertainty in $\hat{\bfx}$ through the {\it{assumed}} probability distribution $\pi(\bfx\mid \bfy)$; An OSUE provides direct uncertainty assessment of $\hat{\bfx}$ via its posited true model through the probability distribution $\pi(\bfx \mid \hat{\bfx})$ that incorporates the prior uncertainty in $\bfx$, forward model, the instrument, and the retrieval algorithm. The distribution $\pi(\bfx \mid \hat{\bfx})$ directly reflects the posterior uncertainty about $\bfx$ given the estimated geophysical quantity $\hat{\bfx}$. It also allows the quantification of the model discrepancy between the FP forward model $\bfF(\cdot)$ and the true physical process. In both standard and probabilistic assessment of retrievals, the FP forward model in the OCO-2 mission is expensive to run and the retrieval algorithm is even more computationally expensive to execute since tens of thousands of evaluations of the FP forward model are required  to solve the optimization problem in the retrieval algorithm. Besides, large-scale Monte Carlo simulations are needed  for probabilistic assessment of retrievals due to the complexity of atmospheric conditions. Thus, direct usage of the forward model makes the large-scale OSUEs computationally prohibitive. 

To facilitate the probabilistic assessment of retrievals via OSUEs, we propose to develop an emulator for the forward model $\bfF(\cdot)$.  Such emulator provides a computationally efficient way to obtain the radiance spectra $\bfy$ and also allows its uncertainty to be further propagated into the retrieval algorithm so that the posterior distribution $\pi( \bfx \mid \hat{\bfx})$ can be obtained. It is clear that probability distribution $\pi(\bfx \mid \hat{\bfx} )$ is different from the probability distribution $\pi(\bfx \mid \bfy)$, since the quantity $\pi( \bfx \mid \hat{\bfx})$ takes into account the uncertainties in the forward model, the instrument, and the retrieval algorithm and allows direct uncertainty assessment when combined with the synthetic truth $\bfx$ in OSUEs. In contrast, one might try to build an emulator to directly approximate the retrieval algorithm $\mathcal{R}(\cdot, \bfF)$ in Figure~\ref{fig: OSUE} by treating $\bfy$ as an input and the retrieval algorithm as a black box with output $\hat{\bfx}$. Such emulator construction allows uncertainty propagation from $\bfy$ to $\hat{\bfx}$, and can be used for standard assessment of retrievals; this approach, however, does not allow uncertainty assessment of the instrument and the retrieval algorithm. It only gives the posterior distribution of $\bfx$ given $\bfy$, which cannot be used for direct uncertainty assessment in $\hat{\bfx}$ with the synthetic truth $\bfx$. In addition, when model discrepancy between the physical process simulated by the computer model and reality is a concern, the emulator for the physical forward model together with ground measurements can be used to model such discrepancy within the Kennedy-O'Hagan framework \citep{Kennedy2001}. These reasons motivate us to develop an emulator for the forward model in the OCO-2 mission. 
 
\subsection{OCO-2 Data}

\begin{figure}[htbp]
\begin{center}
\makebox[\textwidth][c]{ \includegraphics[width=1.0\textwidth, height=0.20\textheight]{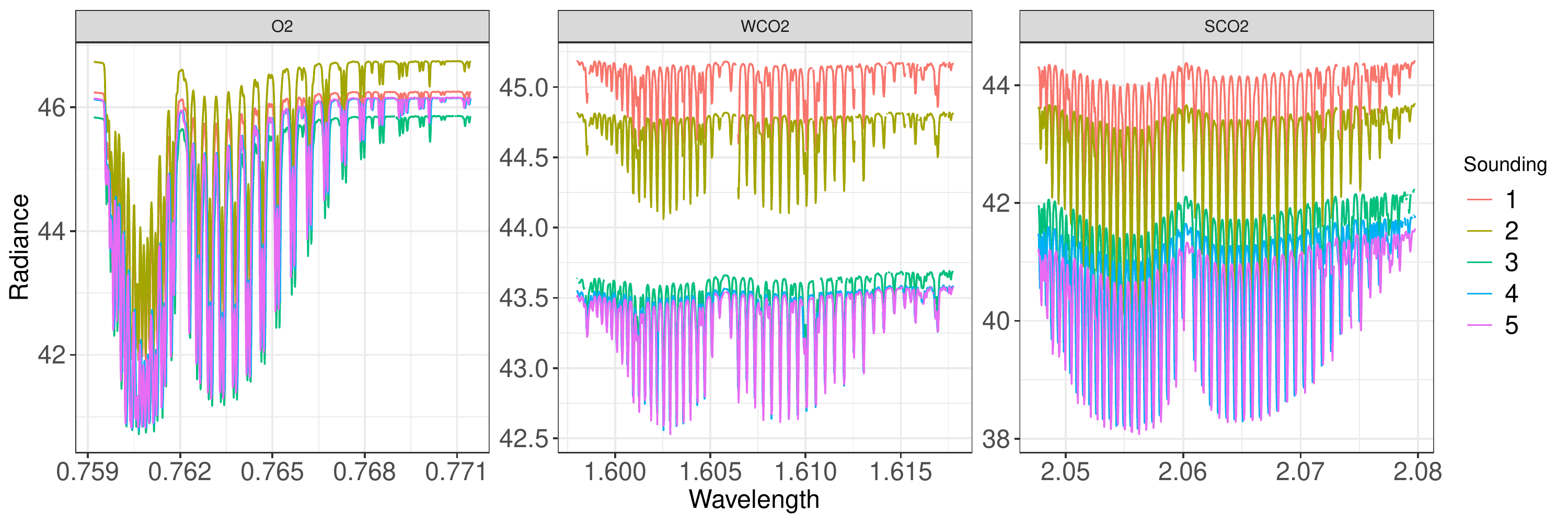}}
\caption{Example of five OCO-2 soundings at three different spectral bands. The log-transformed radiance is plotted against wavelength (in $\mu$m) for each band.}
\label{fig: sounding}
\end{center}
\end{figure}

The OCO-2 instrument carries three imaging grating spectrometers measuring solar radiation reflected from the Earth's surface in the infrared portion of the spectrum. The spectrometers measure the radiation intensity in the three relatively small wavelength bands. These bands, O$_2$-A, weak CO$_2$, strong CO$_2$, are centered near 0.76 $\mu$m, 1.61 $\mu$m, and 2.06 $\mu$m, respectively. In what follows, we will refer to these bands as O$_2$ band, WCO$_2$ band and SCO$_2$ band, respectively. The portion of the radiance vector $\mathbf{y}$ corresponding to each band is a function of different wavelengths with length 1,016. The collection of observed radiances from the three spectral bands at a particular time makes up a \emph{sounding}. Figure~\ref{fig: sounding} shows log-transformed radiances for the three spectral bands from five different OCO-2 soundings. 

The instrument collects eight soundings over its 0.8-degree-wide swath every 0.333 seconds \citep{boesch2017}. The observed radiances result from the interaction between the radiation and the composition of the atmosphere and the Earth's surface along the path from the top of the atmosphere to the surface and back to the satellite. 
 
The emulator developed in this work is based on the FP forward model implemented in Version 7 of the OCO-2 data products \citep{ElderingAMT2017}. The retrieved state vectors, observed radiances, and forward model evaluations are part of the OCO-2 Level 2 diagnostic data products, available at the NASA Goddard Earth Science Data and Information Services Center (GES DISC, \url{https://disc.gsfc.nasa.gov/datasets?page=1\&keywords=OCO-2}). We utilize data that include a state vector $\bfx$ of dimension 62 and augmented with the FP forward model Jacobians, consistent with the configuration used by \citet{ConnorEtAl2016}. The state vector characterizes the CO$_2$ vertical profile, surface pressure, surface albedo, aerosols, temperature and water vapor profile offsets, solar induced fluorescence, and wavelength offsets for the instrument. In developing the emulator, we also include the FP forward model dependence on viewing geometry parameters $\bfb$. These include instrument azimuth angle, instrument zenith angle, solar azimuth angle, and solar zenith angle. The evaluation of the FP forward model and Jacobians for a single sounding takes approximately 40 seconds on a single 2.6 GHz processor that is part of the OCO-2 operational computing cluster.

\section{Model Formulation} \label{sec: Model}
Let $y_k(\bfx, \bfb; \omega)$ be a spectrum radiance at wavelength $\omega$ generated form the FP forward model at input $(\bfx, \bfb)$ for the $k$-th spectral band, where $\bfx$ is a $62$-dimensional vector containing the atmospheric state, and $\bfb$ is a 4-dimensional vector containing viewing geometry. For the O$_2$ band, $\omega$ takes 1,016 different values near 0.76$\mu$m; for the WCO$_2$ band, $\omega$ takes 1,016 different values near 1.61 $\mu$m; for the SCO$_2$ band, $\omega$ takes 1,016 different values near 2.06$\mu$m. Let $\bfy_k(\bfx, \bfb):=(y_k(\bfx, \bfb; \omega_{k,1}), \ldots, y(\bfx, \bfb; \omega_{k,1016}))^\top$ be a sounding consisting of radiances at 1,016 wavelengths for the O$_2$ band ($k=1$), WCO$_2$ band ($k=2$), and SCO$_2$ band ($k=3$).

The goal here is to construct an emulator that can approximate the FP forward model output at arbitrary input settings over the input space. To deal with multivariate output, separability between input space and output has often been assumed to reduce computational burdens \cite[e.g.,][]{Conti2010,BILIONIS2013212,Gu2016, Guillas2018}, that is, the covariance matrix of data can be written as a {Kronecker} product of the input covariance matrix and the output covariance matrix. The separability assumption has been proven quite successful in many real applications when outputs are generated over regularly-spaced grid for different inputs. However, the output in the OCO-2 data is a function of  wavelengths with missing values.  These methods cannot be applied readily in this situation.    

To model such high-dimensional functional output within each band, we adopt a basis-function representation approach similar to the approaches in \cite{Bayarri2007} and \cite{Higdon2008}. Such basis-function representation allows for non-separability between input space and output space \citep[e.g.,][]{Fricker2013} within each band. As there is a wavelength gap between each spectral band, it is natural to assume the separable basis-function representation for each band.  For the $k$-th spectral band, we assume that the functional output $y_k(\bfx, \bfb; \omega)$ has a truncated basis-function representation:
\begin{align} \label{eqn: model for y}
    {y}_k(\mathbf{x}, \mathbf{b}; \omega) = \mu_k(\omega) + \sum_{i=1}^{p_k}\eta_{k,i}(\omega) z_{k,i}(\bfx, \bfb), 
\end{align}
where $\mu_k(\omega)$ is a mean function,  $\eta_{k,i}(\omega)$'s are the basis functions at wavelength $\omega$, and $z_{k,i}(\bfx, \bfb)$'s are the FPCA weights that are assumed to be independent GPs over the input space. Unlike the characteristics of computer model outputs in \cite{Higdon2008}, the OCO-2 outputs are radiances at irregular wavelengths, varying across spatial locations and containing many missing values.  Thus, we cannot construct the basis functions via principal component analysis directly. 

To deal with irregularly-spaced output at varying wavelengths and missing values, we use a functional principal component analysis (FPCA) to construct basis functions. In the FPCA approach, the mean function $\mu_k(\omega)$ and the basis functions $\eta_{k,i}(\omega)$ are estimated as functions of wavelength, which can deal with the irregularly-spaced and/or missing data. Note that \cite{Bayarri2007} use a functional data analysis approach, where a wavelet decomposition is used for irregularly spaced functional outputs. However, the complex modes of variation in the OCO-2 output can only be explained by a prohibitive number basis functions in such a functional representation. Hence we use a FPCA approach, where first a functional representation of the OCO-2 output is obtained using a B-spline basis and then FPCA is applied to the functional data to obtain the main modes of variation through a small number of functional principal components. More specifically, the number of principal components, $p_k$, is chosen such that more than 99\% of the variation is captured. Our numerical results show that the first few functional principal components (at most 3) can explain more than 99\% of the variability for each of the three bands. The detailed description of the FPCA is given in Section~\ref{app: fpca} of the Supplementary Material.

After constructing the mean function $\mu_k(\omega)$ and eigenfunction $\eta_{k,i}(\omega)$, each FPCA weight $z_{k,i}(\bfx, \bfb)$ is then modeled as a mean-zero GP: 
\begin{align} \label{eqn: GP for FPCA weights}
    z_{k,i}(\bfx, \bfb) \sim \mathcal{GP}(0, \sigma^2_{k,i}r((\bfx, \bfb), (\bfx', \bfb'); \bftheta_{k,i})),
\end{align}
where $\sigma^2_{k,i}$ is a variance parameter. $r(\cdot, \cdot)$ is the correlation function over the space on $\bfx$ and $\bfb$ with correlation parameters $\bftheta_{k,i}$. As the dimension of $\bfx$ is $m=62$, $\bftheta_{k,i}$ will contain at least $66=62+4$ parameters in GP emulation. This will cause computational bottlenecks in GP fitting due to the high-dimensionality of the parameter space. To overcome this problem, we project the high-dimensional input space $\mathcal{X}=\{\bfx: \bfx \in \mathbb{R}^m\}$ into a low-dimensional space $\mathcal{S}:=\{\bfs: \bfs=\bfP^\top\bfx\}$ via an $m\times p$ projection matrix $\bfP:=[\bfp_1, \ldots, \bfp_p]$ with $p<m$. Let $\Ell:=(\bfs^\top, \bfb^\top)^\top$ be a vector of input variables in the projected space and viewing geometry. Let $d=\sqrt{\sum_{j=1}^{p+4} (\ell_j-\ell_j')^2/\theta_{kij}^2}$ be the scaled distance between $\Ell$ and $\Ell'$ with $\bftheta_{ki}:=(\theta_{ki1}, \ldots, \theta_{ki(p+4)})^\top$. Then it is easily to recognize that 
\begin{align}
    d = \sqrt{(\bfx - \bfx')^\top \left ( \sum_{j=1}^{p} \frac{\bfp_j\bfp_j^\top}{\theta_{kij}^2}  \right) (\bfx - \bfx') + (\bfb-\bfb')^\top\Lambda (\bfb-\bfb')},
\end{align}
where $\Lambda:=\text{diag}\{1/\theta_{ki(p+1)}^2, \ldots, 1/\theta_{ki(p+4)}^2\}$ is a diagonal matrix with its diagonal entires given by the range parameter for each viewing geometry. Many correlation families can be written as a function of $d$, see Appendix~\ref{app: correlation} for various examples. The form of correlation function is known as the geometrically anisotropic correlation in geostatistics \citep{Zimmerman1993, Cressie1993}. Thus, if $r(\Ell, \Ell'; \bftheta_{k,i})$ is a function of $d$ on $\mathcal{L}\times \mathcal{L}$, it induces a correlation function on $\mathcal{X} \times \mathcal{B}$. This form of correlation function works well in the OCO-2 application, although other forms such as product form are also possible.

The key to maintaining this dual form is to identify the projection matrix $\bfP$. In Section~\ref{sec: AS}, we give a choice of the projection matrix via the \emph{active subspace} method \citep[e.g.,][]{Constantine2014}. With the projection matrix $\bfP$, we can directly work with the independent GPs for $z_{k,i}(\bfs, \bfb)$ on the input space $\mathcal{L}\equiv  \{\bfell: \bfell=(\bfs, \bfb), \mathbf{s} \in \mathcal{S}, \mathbf{b} \in \mathcal{B}\}$ instead of the high-dimensional space $\{ (\bfx, \bfb): \bfx\in \mathcal{X}, \bfb\in \mathcal{B}\}$. In the OCO-2 application, the input dimension is reduced from 66 to 8 via the active subspace approach. The number of model parameters is reduced significantly, resulting in better MCMC mixing and less Metropolis-Hastings steps.   

After performing dimension reduction for both output space and input space, GP modeling needs to be performed based on $10,849$ model runs for the data $\{z_{k,i}(\bfell_j): j=1, \ldots, n\}$ for each $i=1, \ldots, p_k$ and each $k=1,2,3$. For each FPCA weight, the Cholesky decomposition of a $10,849\times 10,849$ covariance matrix is computationally too demanding in the likelihood evaluation. So, we adopt a fast Gaussian process approximation method to reduce the computational challenges in model fitting. Various GP approximations have been proposed in the literature. Here, we choose the nearest neighbor Gaussian process (NNGP), since it has a linear computational cost in the number of model runs, and it possesses good predictive accuracy \citep{Datta2016}. Besides, the NNGP can be viewed as a special instance of the general Vecchia approximation in \cite{Katzfuss2020} that is developed in spatial statistics with competitive performance. The NNGP has an advantage of enabling fully Bayesian inference over other local GP approximations \cite[e.g.,][]{Gramacy2015}. The detailed description of the NNGP is given in Section~\ref{app: NNGP} of the Supplementary Material. We also noticed that the general Vecchia approximation in \cite{Katzfuss2020} could be another promising method in this research since it includes the NNGP as a special case, but its Bayesian formulation and algorithmic development for the OCO-2 application is left for future research.

\subsection{Dimension Reduction via Active Subspace} \label{sec: AS}

The high-dimensional input imposes computational challenges in GP modeling, so we need to perform dimension reduction for the input space. Viewing geometry has known substantial variation in space and time and has a large impact on outputs. We only reduce the dimensionality of the atmospheric state input $\bfx$. As the gradient information for input $\bfx$ is available, we focus on the active subspace approach, although other dimension-reduction approaches such as sufficient dimension reduction (SDR) and effective dimension reduction (EDR) \citep{Cook1994, Cook2009, Li1991} can be alternatives. 

The active subspace approach is closely related to SDR and EDR in regression settings.  SDR and EDR constitute a well-established framework that aims at finding a low-dimensional orthogonal matrix for predictors (or input data) such that the projected data contains as much information as possible on the responses (or output data). In other words, the goal of SDR and EDR is to estimate a projection matrix such that the conditional density (or conditional mean) of the regression function, given a few linear combinations of predictors, is equal to the conditional density (or conditional mean) given all predictors. The active subspace differs from these dimension dimension methods in several aspects: (1) the simulator $\bfF(\cdot)$ is deterministic, so the conditional density of $\bfF(\bfx)$ given $\bfx$ is a Dirac delta function unlike the typical assumption in SDR and EDR; (2) the simulator provides information to compute the gradient with respect to the inputs, which can be exploited in dimension reduction; (3) unlike regression settings, the active subspace is often used to perform UQ studies including emulation, inverse problems, and optimization. For more in-depth discussion, see \cite{Constantine2014}. Below is a brief review of the active subspace method.

The state vector $\bfx$ is assumed to be a Gaussian random vector with known mean $\bfmu_{\bfx}$ and covariance matrix $\bfSigma_{\bfx}=\mathbf{D_{\bfx}R_{\bfx}D_{\bfx}}^\top$, where $\bfD_{\bfx}$ is a diagonal matrix of marginal variance for $\bfx$ and $\bfR_{\bfx}$ is a correlation matrix of $\bfx$. 
The covariance matrix $\bfSigma_{\bfx}$ is the prior covariance matrix that is constructed based on the anticipated global variability of the atmospheric state \citep{boesch2017}. Let the Jacobian of the forward model w.r.t.~$\mathbf{x}$ be $\bfK(\mathbf{x}) \equiv \frac{\partial \bfF(\bfx, \bfb)}{\partial \mathbf{x}}$. Let $\tilde{\bfx}=\mathbf{D}_{\bfx}^{-1} (\bfx - \bfmu_{\bfx})$. Then $\tilde{\bfx} \sim \mathcal{N}_m(\mathbf{0}, \mathbf{R}_{\bfx})$ with $\tilde{\bfx} \in \mathcal{X}$. We define the following data misfit function $f$ according to Equation~\eqref{eqn: forward model}:
\begin{eqnarray*}
f(\tilde{\bfx}) \equiv  [\bfy - \bfF(\bfx, \bfb)]^\top\bfSigma_{\bfeps}^{-1}[\bfy - \bfF(\bfx, \bfb)].
\end{eqnarray*}
Then the derivative of the function $f$ w.r.t.~$\tilde{\mathbf{x}}$ is
\begin{eqnarray*} %\label{eqn: gradient}
\nabla f \equiv \frac{\partial f(\tilde{\bfx})}{\partial \tilde{\bfx}}  = -2 [\bfK(\bfx)\mathbf{D}_{\bfx}]^\top \bfSigma_{\bfeps}^{-1} [\mathbf{y} - \bfF(\bfx, \mathbf{b})]. 
\end{eqnarray*}

Following the procedure in \cite{Constantine2014}, we gave the details of the active subspace in Algorithm~\ref{alg: AS}. We first generate $N$ Monte Carlo samples $\{\bfx_j: j=1, \ldots, N\}$ from the input distribution $\pi(\bfx)$. The number of Monte Carlo samples is set to be large enough so that it is larger than the recommended values in \cite{Constantine2014} and meanwhile these Monte Carlo samples are well spread across the input space. Then the FP forward model is used to generate outputs $\{\bfy_j: j=1, \ldots, N\}$ at these inputs. As the active subspace requires inputs to be drawn from the same input distribution, we standardize them to obtain a new collection of inputs $\{ \tilde{\bfx}_j: j=1, \ldots, N\}$.  It is worth noting that in general the Monte Carlo samples used to find the active subspace do not necessarily need to be used to train the emulator. However, the current work does not distinguish these two situation because standard space-filling designs used for training the emulator can be challenging since the inputs need to have scientific constraints (e.g., satisfying atmospheric conditions over land regions of interest).

\begin{algorithm}[!t]
\caption{Active subspace}
\label{alg: AS} 
\begin{raggedright}
\textbf{Input:} Data $\{ (\bfx_j, \bfy_j): j=1, \ldots, N \}$ and gradient information $\{ \bfK_j: j=1, \ldots, N \}$. \\
 \textbf{Output:} Projection matrix $\bfP$. 
\par\end{raggedright}

 \begin{algorithmic}[1]  
 \State Draw $N$ independent samples $\{\tilde{\bfx}_j: j=1, \ldots, n\}$ from its prior distribution $\pi(\tilde{\bfx})$.
This step is accomplished with samples $\{\bfx_j: j=1, \ldots, N\}$, since $\tilde{\bfx}_j=\mathbf{D}_{\bfx_j}^{-1}(\bfx_j - \bfmu_{\bfx_j})$.

\State Compute the gradient $\nabla f_j \equiv \nabla f(\tilde{\bfx}_j)$ for $j=1, \ldots, N$.

\State Compute eigenvalue decomposition of the $M\times M$ matrix $\bfSigma \equiv 1/N \sum_{j=1}^N (\nabla f_j ) (\nabla f_j)^\top$: 
\begin{eqnarray*}
\bfSigma \equiv \mathbf{\bfW\bfLambda \bfW}^\top,
\end{eqnarray*}
where $\bfW$ is the orthogonal matrix and $\bfLambda=\text{diag}\{\lambda_1, \ldots, \lambda_m\}$ is a diagonal matrix of ordered eigenvalues with $\lambda_1\geq \lambda_2\geq \cdots \geq \lambda_m\geq 0$. The matrix $\bfSigma$ is a Monte Carlo approximation of the sensitivity matrix $E_{\pi(\tilde{\bfx})}[(\nabla_{\tilde{\bfx}} f(\tilde{\bfx}) ) (\nabla_{\tilde{\bfx}} f(\tilde{\bfx}))^\top]$.

\State Let $\mathbf{W} \equiv [ \mathbf{W}_1,\, \mathbf{W}_2]$, and 
$\bfLambda \equiv \text{blockdiag}\{\bfLambda_1, \bfLambda_2\}$,
where $\bfLambda_1$ contains the largest $p$ eigenvalues and $\mathbf{W}_1$ contains the corresponding $p$ eigenvectors. The value of $p$ can be determined based on the gap of eigenvalues and the summation $\sum_{j=1}^p \lambda_j$.

\State Return the project matrix $\bfP: = \bfW_1$. 
\end{algorithmic} 
\end{algorithm}

Once the dimension of the subspace, $p$, is selected, the column space of $\mathbf{P}$ is called an \emph{active subspace}, denoted by $\mathcal{S}\equiv \{\bfs: \bfs\equiv\mathbf{P}^\top \tilde{\bfx}, \, \tilde{\bfx} \in \mathcal{X}\}$, where $\bfs\equiv \bfP^\top\tilde{\bfx}$ is called \emph{active variable}. The function $f$ is most sensitive to the active subspace $\mathcal{S}$, and is almost flat in the space $\mathcal{S}^{\perp}=\{\mathbf{u}: \mathbf{u}\equiv \bfW_2^\top\tilde{\bfx},\, \tilde{\bfx} \in \mathcal{X}\}$ on average.

To select the active subspace, the evaluation of the gradient function $\nabla f$ is required. The construction of the active subspace needs $N$ times evaluation of the gradient $\nabla f$, forward function $\bfF$, and Jacobian $\bfK$, which in total requires $2N$ times evaluation of the forward function $\bfF$ and Jacobian $\bfK$. The constructed active subspace depends on the radiance vector $\bfy$. For different $\bfy$, different active subspaces will be selected. The selected subspace $\mathcal{S}$ will be treated as an input space in Gaussian process modeling in Section~\ref{sec: Emulation}.

\subsection{Emulation Across Different Bands} \label{sec: Emulation}
Let $z_{k,i}(\bfell)$ be the $i$-th FPC weight at input $\bfell$ for the $k$-th band, where $k=1, 2, 3$ correspond to the O$_2$ band, WCO$_2$ band, SCO$_2$ band, respectively. $i=1, \ldots, p_k$ correspond to the FPC weights within the $k$-th band. In the OCO-2 application, the O$_2$ band requires three basis functions (i.e., $p_1=3$), the WCO$_2$ band and SCO2 band require only one basis function, respectively (i.e., $p_2=p_3=1$). Let $q=\sum_{k=1}^3 p_k=5$ be the total number of FPC weights across all the three bands. For notational convenience, let $\bfz(\bfell):=(z_{1,1}(\bfell), z_{1,2}(\bfell), z_{1,3}(\bfell),  z_{2,1}(\bfell), z_{3,1}(\bfell))^\top$ be a vector of $q$ FPC weights at input $\bfell$ corresponding to the O$_2$ band, WCO$_2$ band, and SCO$_2$ band. In what follows, we develop a band-independent (or independent) emulator and a band-separable (or separable) emulator.

\subsubsection{Independent Emulator} \label{sec: independent emulator}
It is clear that one wants to assume independent NNGP models for FPCA weights within each band, since the basis functions are orthogonal. In terms of correlation across different bands, one needs to be very careful since not all spectral radiances are correlated in the OCO-2 application due to the characteristics of the underlying physical processes. One simple way to proceed is to assume an independent covariance structure across different bands. At the same time, we allow different model parameters for each band. This leads to an \emph{independent} emulator with \emph{band-specific} model parameters. That is, we assume the following independent NNGP models across all the FPC weights and spectral bands. More specifically, for the $i$-th FPC weights at the $k$-th band, we assume that 
\begin{align} \label{eqn: IND model}
\begin{split}
z_{k,i}(\cdot) \mid \bfbeta_{k,i}, \sigma^2_{k,i},  \bftheta_{k,i}, \tau^2_{k,i} \,\stackrel{ind}{\sim}\, \text{NNGP}(\bfh^\top(\cdot) \bfbeta_{k,i}, \sigma^2_{k,i}r(\cdot, \cdot, \bftheta_{k,i})), & \\
k=1, \ldots, 3; i=1, \ldots, p_k; &
\end{split}
\end{align} 
where $\bfbeta_{k,i}$ is a vector of fixed basis functions. $\sigma^2_{k,i}$ is the variance parameter. $r(\Ell, \Ell', \bftheta_{k,i}):=c(\Ell, \Ell'; \bftheta_{k,i})+\tau^2_{k,i}I(\Ell=\Ell')$. $c(\cdot, \cdot; \bftheta_{k,i})$ is the correlation function induced by a NNGP with possible choices given in Appendix~\ref{app: correlation}. $\bftheta_{k,i}$ denote correlation parameters. $\tau^2_{k,i}$ is the nugget variance parameter.

Given $n$ model runs with the inputs $\mathcal{L}^O:=\{\Ell_i: i=1, \ldots, n\}$ and corresponding outputs $\bfz=:\{ z_{k,i}(\calL^0): k=1,\ldots, 3; i=1, \ldots, p_k\}$, the marginal likelihood of $\bfz$ is a product of each individual marginal likelihood for each FPC weight. The marginal likelihood of $\bfz_{k,i}$ for the $i$-th FPC weights at the $k$-th band is multivariate normal of the form
\begin{align*}
\bfz_{k,i} \mid \bfbeta_{k,i}, \sigma^2_{k,i}, \bftheta_{i,k}, \tau^2_{k,i} \, \sim \, \mathcal{N}_{n}(\bfH\bfbeta_{k,i}, \sigma^2_{k,i}\bfR_{k,i}),
\end{align*}
where $\bfH:=[\bfh(\Ell_1), \ldots, \bfh(\Ell_n)]^\top$ is an $n$-by-$p$ matrix of fixed basis functions. $\bfR_{k,i}:=\bfC_{k,j} + \tau^2_{k,i}\mathbf{I}$ with $\bfC_{k,i}:=[c(\Ell_i, \Ell_j; \bftheta_{k,i})]_{i,j=1, \ldots, n}$. A standard practice is to integrate out the location-scale parameters with respect to the Jeffreys priors $\pi(\bfbeta_{k,i}, \sigma^2_{k,i}) \propto 1/\sigma^2_{k,i}$ so that the integrated likelihood is used for statistical inference \citep{Berger2001}.

To derive the prediction formula for $\bfz(\cdot)$ at any new input $\Ell_0$. we define the following notations: $\bfbeta:=\{\bfbeta_{k,i}:k=1,2,3; i=1,\ldots, p_k\}$, $\bfsigma^2:=\{\sigma^2_{k,i}:k=1,2,3; i=1,\ldots, p_k\}$, $\bftau^2:=\{\bftau^2_{k,i}:k=1,2,3; i=1,\ldots, p_k\}$, and $\bftheta:=\{\bftheta_{k,i}: k=1,2,3; i=1, \ldots, p_k\}$. Let $\pi(\bfbeta, \bfsigma^2) \propto \prod_{k=1}^3\prod_{i=1}^{p_k} 1/\sigma^2_{k,i}$ be the noninformative priors for all the location-scale parameters. The independence assumption agains allows us to write the the posterior predictive distribution of $\bfz(\Ell_0)$ given $\bfz$ and $\bftheta, \bftau^2$ as a product of independent posterior predictive distributions. In fact, the posterior predictive distribution for the $i$-th FPC weight at the $k$-th band is a student $t$ distribution with $n-p$ degrees of freedom:
\begin{align} \label{eqn: prediction under IND}
z_{k,i}(\Ell_0) \mid \bftheta_{k,i}, \tau^2_{k,i}, \bfz_{k,i}\, \sim \, t(\hat{z}_{k,i}(\Ell_0), \hat{\sigma}^2_{k,i}\hat{r}_{k,i}(\Ell_0), n-p),
\end{align}
with 
\begin{eqnarray*} 
\begin{split}
\hat{z}_{k,i}(\bfell_0) &:= \bfh^\top(\bfell_0) \hat{\bfbeta}_{k,i} + \bfr_{k,i}^\top(\Ell_0) r_{k,j}^{-1}(N(\Ell_0), N(\Ell_0)) \{z_{k,i}({N(\bfell_0)}) - \bfh({N(\bfell_0)}) \hat{\bfbeta}_{k,i}\}, \\
\hat{r}_{k,i}(\bfell_0) &:= r_{k,i}(\bfell_0, \bfell_0) - r_{k,i}^\top(\bfell_0) r_{k,i}^{-1}(N(\Ell_0), N(\Ell_0)) r_{k,i}(\bfell_0) \\
			  &\quad + \biggl \{\bfh(\bfell_0) - \bfh^\top(N(\Ell_0)) r_{k,i}^{-1}(N(\Ell_0), N(\Ell_0)) r_{k,i}(\bfell_0) \biggr\}^\top \biggl\{\bfh^\top(\Ell_0) r_{k,i}^{-1}(N(\Ell_0), N(\Ell_0)) \\
			  &\quad\times \bfh(\Ell_0) \biggr\}^{-1}  \times \biggl\{ \bfh(\bfell_0) - \bfh^\top(N(\Ell_0)) r_{k,i}^{-1}(N(\Ell_0), N(\Ell_0)) r_{k,i}(\bfell_0) \biggr\}, \\
\hat{\sigma}^2_{k,i} &:= (n-p)^{-1}(\bfz_{k,i} - \bfH\hat{\bfbeta}_{k,i})^\top\bfR_{k,i}^{-1}(\bfz_{k,i} - \bfH\hat{\bfbeta}_{k,i}),
\end{split}
\end{eqnarray*}
where $\hat{\bfbeta}_{k,i}:=(\bfH^\top \bfR_{k,i}^{-1} \bfH)^{-1} \bfH^\top \bfR_{k,i}^{-1} \bfz_{k,i}$ is the generalized least squares estimator of $\bfbeta_{k,i}$. $\bfr_{k,i}(\Ell_0) := r(N(\bfell_0), \bfell_0; \bftheta_{k,i})$ is a $t$-by-1 vector of cross-correlations between the neighbors of input $\Ell_0$ and $\Ell_0$.  $r_{k,i}({N(\bfell_0), N(\bfell_0)})$ is a $t$-by-$t$ correlation matrix of neighbors of input $\Ell_0$. $z_{k,i}(N(\Ell_0))$ is a $t$-by-1 vector of output at the neighbors of input $\Ell_0$. $\bfh(N(\Ell_0))$ is a $t$-by-$p$ matrix of fixed basis function matrix evaluated at the neighbors of input $\Ell_0$. This predictive distribution takes into account the parameter uncertainty due to estimation of the location-scale parameters $\bfbeta, \sigma^2$. It is only a function of correlation parameters and nugget parameter. Once the correlation parameters and nugget parameter are known, this convenient form allows efficient downstream computation based on the predictive distribution. 

To estimate correlation parameters $\bftheta_{k,i}$ and $\tau^2_{k,i}$, we recommend working with the integrated posterior of these parameters given the data $\bfz$ after integrating out the location-scale parameters. It has been recognized that the MCMC algorithm based on latent spatial variables in the NNGP model \citep{Datta2016} suffers from MCMC convergence issues for a high-dimensional model \citep{Finley2019}. To overcome such issues, we integrate out all the latent spatial variables as well as the location-scale parameters, and derive the marginal posterior of correlation parameters and nugget parameters given the data.    

More specifically, we assign the noninformative priors of the form: $\pi(\lambda) = 1/(1+\lambda^2)$ for $\lambda>0$, where $\lambda$ represents any scalar in $\{\bftheta_{k,i}, \tau^2_{k,i}; k=1,2,3; i=1,\ldots, p_k\}$. Let $\pi(\bftheta_{k,i}, \tau^2_{k,i})$ denote the prior distribution for $\bftheta_{k,i}$ and $\tau^2_{k,i}$. The posterior distribution of correlation parameters and nugget parameters is also a product of each posterior distribution for each FPC weights. The posterior distribution of $\bftheta_{k,i}$ and $\tau^2_{k,i}$ is:
\begin{align} \label{eqn: posterior under IND}
\pi(\bftheta_{k,i}, \tau^2_{k,i} \mid \bfz_{k,i}) \propto \pi(\bftheta_{k,i}) |\bfR_{k,i}|^{-1/2} |\bfH^\top\bfR_{k,i}^{-1} \bfH|^{-1/2} |\bfz_{k,i}^\top \bfG_{k,i}\bfz_{k,i}|^{-(n-p)/2}, 
\end{align}
where $\bfG_{k,i}:=\bfR_{k,i}^{-1} - \bfR_{k,i}^{-1} \bfH(\bfH^\top \bfR_{k,i}^{-1} \bfH)^{-1} \bfH^\top \bfR_{k,i}^{-1}$. The computation of this posterior involves utilization of sparse matrices as illustrated in Appendix~\ref{app: computation}. This posterior distribution allows both empirical Bayes (maximum a posteriori) estimation  and fully Bayesian inference via standard MCMC algorithms, where log-normal distributions are used as proposal distributions in the Metropolis-Hastings steps. This posterior distribution allows efficient posterior sampling compared to the sequential updating scheme in the NNGP model \citep{Datta2016} as confirmed by our application. 

In the UQ community, an empirical Bayes estimation approach has often been used for Gaussian process emulation because fully Bayesian inference can be too expensive to estimate correlation parameters in most applications. Our formulation of the NNGP model facilitates both the empirical Bayes and fully Bayesian approaches. In this application, we choose to deploy the fully Bayesian inference via MCMC algorithms because this emulator will be used in the inverse UQ part of the OSUEs, where the inverse problem will be solved with an MCMC algorithm as opposed to the OE operated in the OCO-2 mission.   

\subsubsection{Separable Emulator} \label{sec: SEP}
 The physics of molecular absorption of CO$_2$ suggest that the radiances in WCO$_2$ and SCO$_2$ bands respond similarly to the same profile of atmospheric CO$_2$. This physical correlation would be ignored via the \emph{independent} emulator in Section~\ref{sec: independent emulator}. To overcome such limitation, we also develop an emulator with a separable covariance model across different bands. The resulting emulator will be called a \emph{separable} emulator. It is worth mentioning that for both independent emulator and separable emulator, the covariance structure between radiance output and input space within each band is nonseparable due to the basis-function representation, while the covariance structure across different bands is separable.

To account the cross-correlation among FPC weights across different bands, we assume a separable structure as in \cite{Conti2010}, that is, the correlation function of FPC weights across all the bands is a product of input correlation function and an unknown correlation matrix that models the cross-correlations between FPC weights at different bands. In what follows, we give the model specification in terms of all the $q$ FPC weights for notational convenience, however, in practice, we only implement the separable emulator for the FPC weights from the WCO$_2$ band and SCO$_2$ bands.

Specifically, a $q$-variate Gaussian process with a separable covariance structure across different bands would assume the following form
\begin{align} \label{eqn: separable emulation}
\bfz(\cdot) \mid \bfbeta, \bfSigma, \bftheta, \tau^2 \sim \mathcal{GP}(\bfh(\cdot)^\top \bfbeta,\, r(\cdot, \cdot) \bfSigma).
\end{align}
where $\bfh(\bfx)$ is a vector of $p$ basis functions at input $\bfx$. $\bfbeta$ is the $p\times q$ matrix of unknown regression coefficients. $r(\Ell, \Ell')=c(\Ell, \Ell'; \bftheta) + \tau^2I(\Ell=\Ell')$, where $c(\cdot, \cdot; \bftheta)$ is the input correlation function with parameters $\bftheta$. $\tau^2$ is the nugget parameter. $\bfSigma$ is the unknown covariance matrix that takes into account cross-covariance among $q$ output variables at any input. 

Let $\bfZ:=[\bfz(\bfell), \cdots, \bfz(\bfell_n)]^\top$ be the $n\times q$ matrix of output. The marginal likelihood of $\bfZ$ is a matrix normal distribution of the form 
\begin{align*}
\bfZ \mid \bfbeta, \bfSigma, \bftheta, \tau^2 \sim \mathcal{MN}_{n,q}(\bfH \bfbeta, \bfR, \bfSigma),
\end{align*}
where $\bfH:=[\bfh(\bfell_1), \ldots, \bfh(\bfell_n)]$ is the $n\times p$ matrix of basis functions. $\bfR:=\bfC + \tau^2\mathbf{I}$, where $\bfC:=[c(\bfell_i, \bfell_j; \bftheta)]_{i,j=1, \ldots, n}$ is the $n\times n$ correlation matrix on the set of inputs $\mathcal{L}^O$. As the number of training samples is large, we again assume the correlation function $c(\cdot, \cdot)$ to be induced by a NNGP prior to overcome the computational bottlenecks.

Given the model~\eqref{eqn: separable emulation} and $n\geq p+q$ so that all ensuing posteriors are proper, the posterior predictive distribution of $\bfz(\bfell_0)$ given $\bftheta, \tau^2$ and $\bfZ$ is a $q$-variate $t$ process ($\mathcal{TP}$) such that the distribution of an arbitrary collection of vectors is a matrix-variate distribution. The posterior predictive distribution is a $q$-variate $t$-distribution of the form,
\begin{align}\label{eqn: prediction under SEP}
z(\bfell_0) \mid \bftheta, \tau^2, \bfZ \sim \mathcal{MVT}(\hat{z}(\bfell_0), \hat{r}(\bfell_0) \hat{\bfSigma}_{gls}, n-p),
\end{align}
with $n-p$ degrees of freedom, in which 
\begin{align*}
\hat{z}(\bfell_0) &:= \bfh^\top(\bfell_0) \hat{\bfbeta}_{gls}  + r(\bfell_0) r^{-1}(N(\Ell_0), N(\Ell_0)) \{z(N(\Ell_0)) - \bfh(N(\Ell_0))\hat{\bfbeta}_{gls}\}, \\
\hat{r}(\bfell_0)&:= r(\bfell_0, \bfell_0) - r^\top(\bfell_0) r^{-1}(N(\Ell_0), N(\Ell_0)) r(\bfell_0) \\
			  &\quad + \biggl \{\bfh(\bfell_0) - \bfh^\top(N(\Ell_0)) r^{-1}(N(\Ell_0), N(\Ell_0)) r(\bfell_0) \biggr\}^\top \biggl\{\bfh^\top(\Ell_0) r^{-1}(N(\Ell_0), N(\Ell_0)) \bfh(\Ell_0) \biggr\}^{-1} \\
			  &\quad \times \biggl\{ \bfh(\bfell_0) - \bfh^\top(N(\Ell_0)) r^{-1}(N(\Ell_0), N(\Ell_0)) r(\bfell_0) \biggr\}, \\
\hat{\bfSigma}_{gls} & := (n-p)^{-1} (\bfZ - \bfH\hat{\bfbeta}_{gls})^\top \bfR^{-1} (\bfZ - \bfH\hat{\bfbeta}_{gls}),
\end{align*}
where $\hat{\bfbeta}_{gls}:= (\bfH^\top \bfR^{-1} \bfH)^{-1} \bfH^\top \bfR^{-1} \bfZ$ is the generalized least squares estimator of $\bfbeta$. $z(N(\Ell_0))$ is a vector of nearest neighbor outputs for the input $\Ell_0$. $\bfh(N(\Ell_0))$ is a $t$-by-$p$ matrix of fixed basis functions evaluated at the nearest neighbors of input $\Ell_0$. $r(N(\Ell_0),$ $N(\Ell_0))$ is the input correlation matrix of nearest neighbors for $\Ell_0$.  $r(\bfell_0):=r(N(\Ell_0), \Ell_0)$ is a vector of cross-correlations between the nearest neighbors of input $\bfell_0$ and the input $\Ell_0$.

To perform Bayesian inference, we follow \cite{Berger2001} and assign a non-informative prior for $\bfbeta, \bfSigma$ when the parameters $\bftheta$ and $\tau^2$ are given. Similar to Section~\ref{sec: independent emulator}, we assume the Jeffreys prior, i.e., $\pi(\bfbeta, \bfSigma \mid \bftheta, \tau^2)  \propto |\bfSigma|^{-(q+1)/2}$. The priors on $\bftheta$ and $\tau^2$ follow standard prior specification as in Section~\ref{sec: independent emulator}.  Integrating out $\bfbeta, \bfSigma$ yields the following posterior distribution:
\begin{align} \label{eqn: posterior under SEP}
\pi(\bftheta \mid \bfZ) \propto \pi(\bftheta) |\bfR|^{-q/2} |\bfH^\top\bfR^{-1} \bfH|^{-q/2} |\bfZ^\top \bfG\bfZ|^{-(n-p)/2}, 
\end{align}
where $\bfG:=\bfR^{-1} - \bfR^{-1} \bfH(\bfH^\top \bfR^{-1} \bfH)^{-1} \bfH^\top \bfR^{-1}$. This posterior distribution allows fully Bayesian inference via standard MCMC algorithms. Efficient computations utilizing sparse matrices are detailed in Appendix~\ref{app: computation}. 

An interesting note is that we have formulated an upgraded version of the classic separable model \citep{Conti2010} using a NNGP model, based on which legitimate Bayesian inference is permitted. The resulting predictive distribution is a function of correlation parameters and nugget parameter. Besides, the posterior distribution in our formulation allows both empirical Bayes estimation and fully Bayesian estimation. This is in parallel to the posterior based on a typical Gaussian process model. Our development can effectively deal with large-scale computer experiments in the input space compared to the classic separable model.

\subsubsection{Irrelevance of Cross-correlations Across Different Bands}
Given the independent emulator and the separable emulator, one may wonder which one performs better. This question has been illustrated in several previous works \citep[e.g.,][]{Fricker2013, Mak2018}. For instance, \cite{Fricker2013} found that the independent emulator gives better predictive performance in terms of RMSPE than the separable emulator in several numerical examples. In contrast, \cite{Mak2018} demonstrate that the separable emulator with a positive-definite cross-covariance matrix gives better predictive performance than the separable emulator with a diagonal cross-covariance matrix in terms of \emph{joint} confidence regions theoretically under certain conditions in their Theorem 3.1. Their application on turbulent flows also confirms such benefit. This is one advantage of using a separable emulator. The following theorem provides an answer to compare one instance of the independent emulator (with the same input correlation) with the separable emulator. The proof of this theorem is given in Appendix~\ref{app: theorem}. 

\begin{theorem} \label{thm: irrelevance}
Suppose that $\bfz(\cdot):=(z_1(\cdot), \ldots, z_q(\cdot))^\top$ is a $q$-variate process that is observed at $n$ inputs with the $n$-by-$q$ output matrix $\bfZ:=(\bfz(\Ell_1), \ldots, \bfz(\Ell_q))^\top$. Let 
$
\mathcal{M}_0: \bfz(\cdot) \mid \bfbeta, \bfSigma, \bftheta, \tau^2 \sim \mathcal{GP}(\bfh(\cdot)^\top \bfbeta,\, r(\cdot, \cdot) \bfT)
$
be the independent model with the same input correlation parameters. Let 
$
\mathcal{M}_1: \bfz(\cdot) \mid \bfbeta, \bfSigma, \bftheta, \tau^2 \sim \mathcal{GP}(\bfh(\cdot)^\top \bfbeta,\, r(\cdot, \cdot) \bfSigma)
$
be the separable model. Assume that  $\bfT = \text{diag}(\bfSigma)$. Suppose that the Jeffreys prior is used for the location-scale parameters in $\mathcal{M}_0$ and $\mathcal{M}_1$, where $r(\cdot, \cdot)$ is the input correlation function with correlation parameters $\bftheta$. $\tau^2$ denotes the nugget parameter in the correlation function $r(\cdot, \cdot)$. Then it follows that $\bfz(\Ell_0)$ given the parameters $\bftheta, \tau^2$ has the same marginal posterior predictive distributions with identical predictive mean and  identical predictive variance under $\mathcal{M}_0$ and $\mathcal{M}_1$, that is,
\begin{align*} \vspace{-1cm}
E[\bfz(\bfx_0) \mid \bftheta, \tau^2, \bfZ, \mathcal{M}_0] &= E[\bfz(\bfx_0) \mid \bftheta, \tau^2, \bfZ, \mathcal{M}_1], \\
\text{var}[z_k(\bfx_0) \mid \bftheta, \tau^2, \bfZ, \mathcal{M}_0] &= \text{var}[z_k(\bfx_0) \mid \bftheta, \tau^2, \bfZ, \mathcal{M}_1], k=1, \ldots, q.
\end{align*}
\end{theorem}

Theorem~\ref{thm: irrelevance} implies that any pointwise predictive measures based on the marginal posterior predictive distributions should be the same under the independent emulator with the same input correlation and the separable emulator. Note that the independent emulator in Theorem~\ref{thm: irrelevance} is a special case of the independent emulator in Section~\ref{sec: independent emulator}, since the latter allows different input correlation parameters for different outputs. Numerical comparison of the proposed independent emulator with the separable emulator is given in Section~\ref{sec: IND and SEP} and Section~\ref{app: numeric} of the Supplementary Material.

Although the separable emulator provides a way to capture the cross-covariance across different spectral bands, other approaches might be used to develop an emulator that can also account for the cross-covariance. For instance, one may develop an emulator with a nonseparable covariance structure across different bands based on \emph{convolution methods} \citep{Higdon2002} and \emph{linear model of coregionalization} \citep{Journel1978} in spatial statistics, see \cite{Fricker2013} for an overview from the perspective of computer model emulation. However, how to develop an efficient emulator with a nonseparable covariance structure across different bands is still an open question due to the computational challenges in the OCO-2 application.

\section{Application to the FP Forward Model} \label{sec: results}

In this section, we implement the emulators in Section~\ref{sec: Model} with the NASA OCO-2 data. As the input space is of high dimensionality, a large number of samples from the input space are required, so that the input space is well explored by these inputs. In addition we must ensure that the samples from the input space span realistic conditions that OCO-2 could plausibly encounter. Moreover, the number of samples needs to be large enough in order to select the active subspace that not only well represent the input space but also reduce the Monte Carlo approximation error. Therefore we have assembled a sample of $10,894$ state vectors and FP forward model evaluations from the full OCO-2 record of land retrievals for February 2015. This collection has also been used by the OCO-2 team for retrieval error analysis, following \citet{ConnorEtAl2016}, and gives suitably global coverage for constructing the emulator. {The viewing geometry parameters are determined from the location and time of an individual sounding, along with the satellite's orientation.}

\begin{figure}[hbt!]
\begin{center}
\makebox[\textwidth][c]{ \includegraphics[width=1.0\textwidth, height=0.30\textheight]{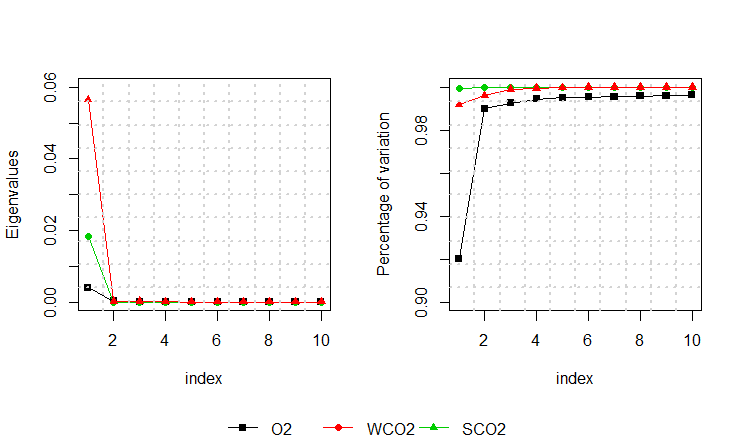}}
\caption{Functional PCAs for the radiance vector over O$_2$, WCO$_2$, SCO$_2$ bands. The left panel shows the eigenvalues. The right panel shows the percentage of cumulative variations.}
\label{fig: fPCA eigen}
\end{center}
\end{figure}

\begin{figure}[hbt!]
  \centering
\makebox[\textwidth][c]{ \includegraphics[width=1.0\linewidth, height=0.25\textheight]{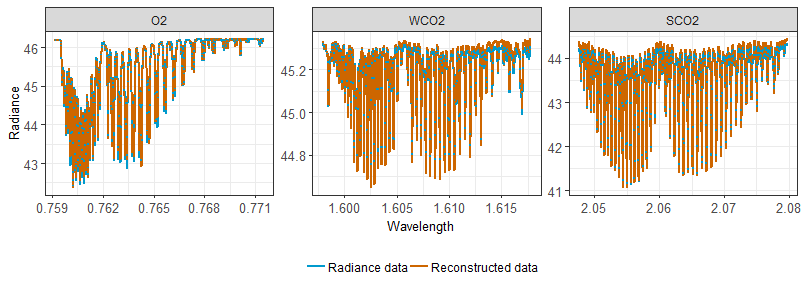}}
\caption{The FP forward model simulated radiances and the corresponding reconstructed radiances using the FPCA over the three bands for a sounding.}
\label{fig: fPCA radiance 15}
\end{figure}

%\subsection{Results for FPCA}
To represent the functional output via a basis function expansion, we perform FPCA on the log-transformed radiances in separately for each of the three bands: O$_2$ band,  WCO$_2$ band and  SCO$_2$ band, since the wavelengths of these three bands do not overlap. The B-spline basis functions are used to convert the simulated radiance data to a functional form of wavelength using the \textsf{R} package \texttt{fda} \citep{fda}. {To preserve the sharp changes of the radiances in this functional form, $500$ B-spline basis functions with equidistant knots over the range of wavelengths in each of the three bands are used}. The eigenvalues in Figure~\ref{fig: fPCA eigen} show that the first three principal components for the O$_2$ band can explain more than 99\% of variance in the output. For the WCO$_2$ and SCO$_2$ bands, the first principal component can explain more than $99 \% $ of variability. Figure~\ref{fig: fPCA radiance 15} shows the true radiances and the reconstructed radiances via the FPCA approach for a sounding. We can see that the reconstructed radiances, using the first three functional principal components for O$_2$ band and one component each for WCO$_2$ and SCO$_2$ band, are very close to the true radiances. This indicates that the main modes of variation of the OCO-2 radiance vectors can be explained through a very small number of functional principal components. 

\begin{figure}[hbt!]
%\begin{center}
\makebox[\textwidth][c]{ \includegraphics[width=1.0\textwidth, height=0.2\textheight]{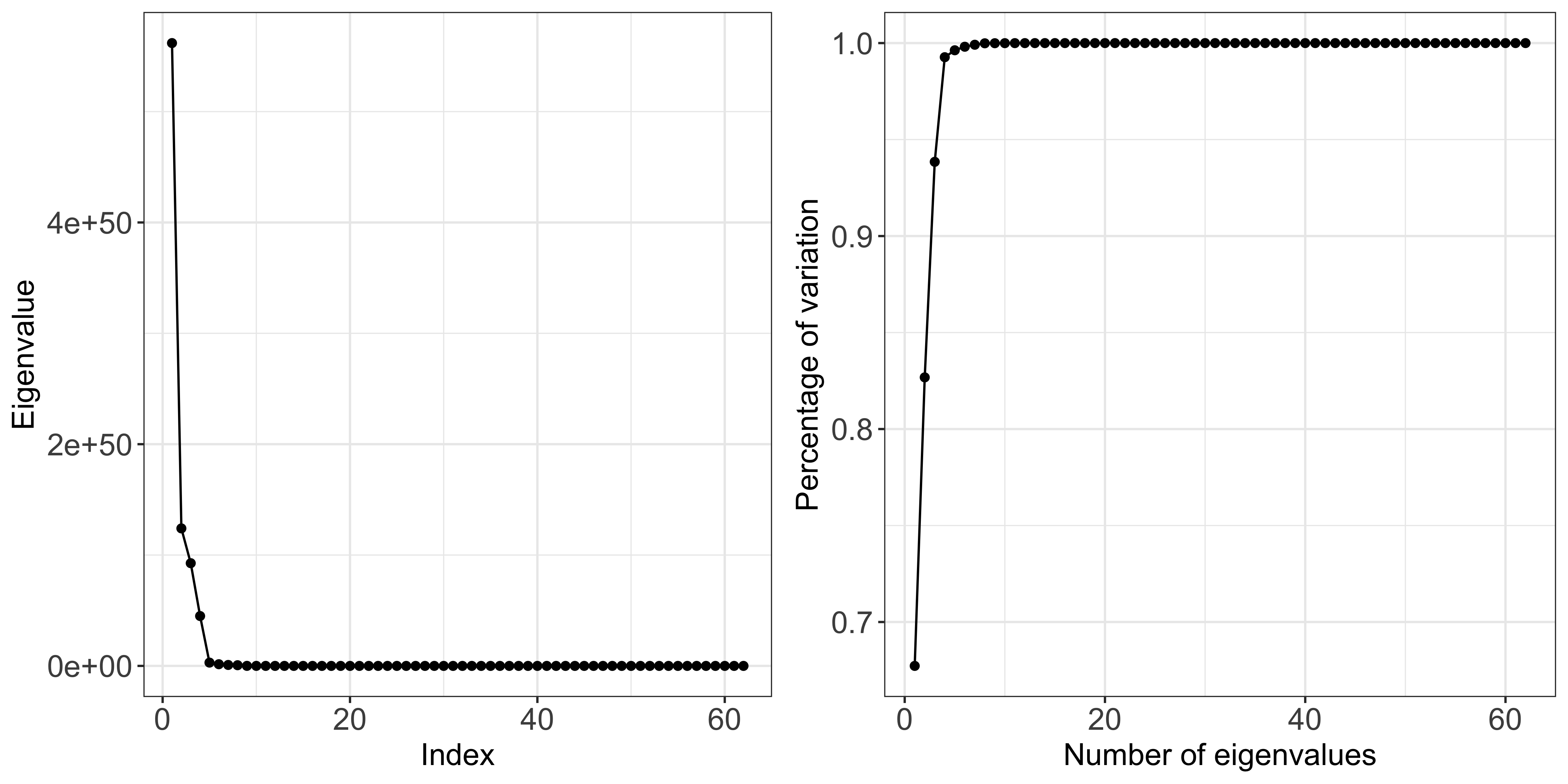}}
\caption{Diagnostics in the active subspace method. The left panel shows the eigenvalues. The right panel shows the percentage of cumulative variations.}
\label{fig: AS for inputs}
%\end{center}
\end{figure}

For input space dimension reduction, Figure~\ref{fig: AS for inputs} shows that four active variables are enough to capture above 95\% of the variability via the active subspace approach. There is a sharp change in the eigenvalues starting from the fourth eigenvalue. This suggests that four active variables can be used with theoretical justification given in \cite{Constantine2014}. It is expected that the active subspace approach results in a smaller number of important variables, since this approach uses the gradient information of the forward model to select active variables that have most impact on the outputs. As a comparison, we also use principal component analysis (PCA) to reduce the dimensionality of input space, with its results shown in Section~\ref{app: PCA} of the Supplementary Material. We found that the active subspace method is a better alternative dimension reduction approach than the PCA approach in the OCO-2 application, since it incorporates gradient information of the FP forward model to perform dimension reduction.

\subsection{Model Comparison with PCA and LaGP} \label{sec: comparison with PCA and LaGP}
In this section, we evaluate the performance of independent emulator under the PCA and active subspace methods. As a comparison, we also use the local approximate Gaussian process approximation \citep{Gramacy2015}. This method will be referred to as LaGP hereafter. For both NNGP and LaGP, the predictive performance is evaluated for $n^*=1,000$ randomly held-out inputs based on PCA and active subspace. For the proposed emulators based on NNGP and the LaGP, we assume a constant mean function for this application. 

To measure the predictive performance, we computed three criteria: root-mean-squared-prediction error (RMSPE), coverage probability of 95\% credible interval ($P_{CI}(95\%)$), and continuous-rank-probability score \citep[CRPS;][]{Gneiting2007}. The RMSPE measures the discrepancy between the forward model output and predictive mean. The coverage probability of the 95\% credible interval measures the percentage of 95\% credible intervals covering the forward model output. The CRPS measures the uncertainties in the predictive distribution.

In the cross-validation study, the NNGP is implemented with 20 nearest neighbors in \textsc{MATLAB} since it works quite well in the application. The reference set is chosen to be the set of training inputs $\mathcal{L}^O$ whose elements are ordered by the first active variable. Other types of ordering strategies give similar prediction results. We found that the anisotropic Mat\'ern correlation with smoothness parameter $\nu=2.5$ works well in the application. We also found that increasing the number of nearest neighbors can significantly increase the computing time and the improvement in terms of prediction is negligible compared to the efforts for computation. For instance, with 60 neighbors, the computing time for one evaluation of the posterior distribution in Equation~\eqref{eqn: posterior under IND} is increased roughly by a factor of six compared to that with 20 nearest neighbors on the same computer. This can significantly slow the MCMC algorithm since each iteration requires nine evaluations of the posterior distribution based on the active subspace approach. The corresponding improvement in terms of RMSPE is less than 30\% when the number of nearest neighbors is increased from 20 to 60 for all the bands. For the MCMC algorithm, the posterior samples for each parameter start to show convergence around 500 iterations in both the independent emulator and the separable emulator. We compute all the posterior summaries based on 3000 samples after the burn in period for each parameter. This finding shows a significant improvement of the proposed Bayesian inference approach compared to the sequential updating scheme in \cite{Datta2016}, where tens of thousands of iterations are typically required. The LaGP is implemented with \textsf{R} package \texttt{laGP} \citep{Gramacy2016}. It involves approximating the predictive equations at a particular generic location, via a subset of the data, where the sub-design is (primarily) comprised of a neighborhood close to the location. These design points are chosen via a sequential approach using different criteria, viz. active learning Cohn (ALC), minimum mean square prediction error (MSPE) and nearest neighbors (NN).

In terms of the choice between PCA and active subspace, the results in Table~\ref{table: predictive measures (PCA)} and Table~\ref{table: predictive measures (AS)} show that the active subspace approach gives a slightly better predictive performance than the PCA approach in terms of RMSPE and CRPS for O$_2$ and WCO$_2$ bands, while the active subspace approach gives a slightly worse result than the PCA approach for the SCO$_2$ band. However, the emulator with the PCA approach requires about five more parameters than the emulator with the active subspace approach. This reduces computing time significantly when fully Bayesian inference is performed in the NNGP. 

\begin{table}[hbt!]
\centering
\normalsize
   \caption{Numerical comparison of held-out sets based on NNGP and LaGP using the PCA approach. The predictions are obtained at $n^*=1,000$ new input values. For each spectrum band, we compared RMSPE, $P_{CI}{(95\%)}$, and CRPS. The results are reported for both functional principal scores and radiance data.}
  {\resizebox{1.0\textwidth}{!}{%
  \setlength{\tabcolsep}{2em}
   \begin{tabular}{l c c c c c c} 
   \toprule %\noalign{\vskip 1.5pt}
& \multicolumn{3}{c}{NNGP}  & \multicolumn{3}{c}{LaGP} \\ \noalign{\vskip .5pt} \cmidrule(lr){2-4} \cmidrule(lr){5-7} \noalign{\vskip .5pt} 
 & RMSPE & $P_{CI}{(95\%)}$  & CRPS & RMSPE & $P_{CI}{(95\%)}$  & CRPS \\ \noalign{\vskip .5pt} \hline \noalign{\vskip .5pt} 
 \multicolumn{7}{c}{Results for each functional principal score} \\  \noalign{\vskip .5pt}   \hline \noalign{\vskip .5pt}
O$_2$ (PC 1) & 0.0232  &  0.953  &  0.0126  &  0.0256 & 0.875 & 0.0146 \\ \noalign{\vskip .5pt} 
O$_2$ (PC 2) &  0.0103  &  0.954 &   0.0057   & 0.0113 & 0.925 & 0.0062 \\ \noalign{\vskip .5pt} 
O$_2$ (PC 3) &  0.0019  &  0.950  &  0.0010   & 0.0021 & 0.914 & 0.0011\\ \noalign{\vskip .5pt} 
WCO$_2$ (PC 1) &  0.0350 &   0.955 &   0.0191   & 0.0416 & 0.897 &  0.0233\\ \noalign{\vskip .5pt} 
SCO$_2$ (PC 1) & 0.0597  &  0.950  &  0.0319  & 0.0700 & 0.907 & 0.0393 \\ \noalign{\vskip .5pt} \hline %\noalign{\vskip 1pt}
\multicolumn{7}{c}{Results for radiances at each spectrum band} \\ \noalign{\vskip .5pt} \hline \noalign{\vskip .5pt}
O$_2$  &  0.2412 & 0.9197 & 0.2309    & 0.2636 &--- &--- \\ \noalign{\vskip .5pt} 
WCO$_2$  & 0.2493 & 0.9537 & 0.2367 & 0.2966 &--- &--- \\ \noalign{\vskip .5pt} 
SCO$_2$  & 0.3562 & 0.9315 & 0.3198  & 0.4104 &--- &--- \\
%\noalign{\vskip 1.5pt} 
\bottomrule
   \end{tabular}%
   }}
   \label{table: predictive measures (PCA)}
\end{table}

\begin{table}[hbt!]
\centering
\normalsize
   \caption{Numerical comparison of held-out sets based on NNGP and LaGP using the active subspace approach. The predictions are obtained at $n^*=1,000$ new input values. For each spectrum band, we compared RMSPE, $P_{CI}{(95\%)}$, and CRPS. The results are reported for both functional principal scores and radiance data.}
  {\resizebox{1.0\textwidth}{!}{%
  \setlength{\tabcolsep}{2em}
   \begin{tabular}{l c c c c c c} 
   \toprule %\noalign{\vskip 2.5pt}
& \multicolumn{3}{c}{NNGP}  & \multicolumn{3}{c}{LaGP} \\ \noalign{\vskip .5pt} \cmidrule(lr){2-4} \cmidrule(lr){5-7} \noalign{\vskip .5pt} 
 & RMSPE & $P_{CI}{(95\%)}$  & CRPS & RMSPE & $P_{CI}{(95\%)}$  & CRPS \\ \noalign{\vskip .5pt} \hline \noalign{\vskip .5pt} 
 \multicolumn{7}{c}{Results for each functional principal score} \\  \noalign{\vskip .5pt}   \hline \noalign{\vskip .5pt}
O$_2$ (PC 1) & 0.0202 &   0.950  &  0.0111 & 0.0208 & 0.928 & 0.0116 \\ \noalign{\vskip .5pt} 
O$_2$ (PC 2) & 0.0099  &  0.952  &  0.0054 & 0.0100 & 0.934 & 0.0054 \\ \noalign{\vskip .5pt} 
O$_2$ (PC 3) & 0.0022  &  0.951  &  0.0012 & 0.0023 & 0.921 & 0.0012\\ \noalign{\vskip .5pt} 
WCO$_2$ (PC 1) & 0.0341 &   0.952  &  0.0180 & 0.0357 & 0.912 & 0.0191 \\ \noalign{\vskip .5pt} 
SCO$_2$ (PC 1) & 0.0700  &  0.945  &  0.0375 & 0.0709 & 0.924 & 0.0383 \\ \noalign{\vskip .5pt} \hline \noalign{\vskip .5pt}
\multicolumn{7}{c}{Results for radiances at each spectrum band} \\ \noalign{\vskip .5pt} \hline \noalign{\vskip .5pt}
O$_2$             &  0.2114 & 0.898 & 0.1910 & 0.2220 &--- &--- \\ \noalign{\vskip .5pt} 
WCO$_2$  & 0.2429 & 0.950 & 0.2095 & 0.2544 &--- &--- \\ \noalign{\vskip .5pt} 
SCO$_2$  & 0.4103 & 0.932 & 0.3780 & 0.4150 &--- &--- \\
%\noalign{\vskip 1.5pt} 
\bottomrule
   \end{tabular}%
   }}
   \label{table: predictive measures (AS)}
\end{table}  

To compare NNGP and LaGP, we fit the LaGP emulator using all the three suggested design criteria. As expected the NN method gives the worst performance whereas the ALC and MSPE methods give comparable results. The maximum number of neighbors for each prediction point is kept at 20, which is the same as NNGP. Such choice is to enable a fair comparison, since they make inference (estimation and prediction) based on chosen nearest neighbors. The prediction results for the validation data set are reported in Table~\ref{table: predictive measures (PCA)} and~\ref{table: predictive measures (AS)} for input dimension reduction approaches based on PCA and active subspace, respectively. We can see from these results that RMSPE from NNGP are slightly smaller than those from LaGP. In terms of uncertainty quantification, NNGP gives better results than LaGP in terms of $P_{CI}{(95\%)}$ and CRPS. Note that the coverage probability column is left blank for the LaGP results at the original radiance scale. The \textsf{R} package output for LaGP only provides the mean prediction and the standard error but not the posterior samples. Hence it is difficult to compute the coverage probability for the original radiance data.

It is worth noting that the NNGP gives better results then LaGP for all the three bands in terms of uncertainty quantification. This is possibly due to the following two reasons: (1) the NNGP uses a geometrically anisotropic Mat\'ern correlation function with the smoothness parameter $\nu=2.5$ while the LaGP uses a product of squared-exponential correlation functions; (2) fully Bayesian inference in NNGP allows better uncertainty quantification. The NNGP emulator is much more expensive to fit when compared to LaGP, since the NNGP uses fully Bayesian inference. Moreover, the LaGP is primarily built for fast prediction at some pre-specified input configurations. In the OCO-2 application, the primary objective is to construct an emulator to later take the place of the FP forward model in an entire OSUE. For instance, the prediction from the emulator is needed multiple times in the retrieval algorithm in order to estimate the atmospheric state vectors in the inverse UQ part of the OSUE. Thus, the corresponding input configurations for these predictions are not known beforehand, making the LaGP emulator not applicable in OSUEs. The NNGP is a process based model and hence prediction can be obtained at any arbitrary input configuration from the input space. This flexibility gives another justification for an NNGP emulator instead of a LaGP emulator for OSUEs in the OCO-2 application. 

\subsection{Comparison Between Independent Emulator and Separable Emulator} \label{sec: IND and SEP}
The Independent Emulator assumes an independent covariance structure across different bands. This assumption ignores strong correlation between the WCO$_2$ band and the SCO$_2$ band, which may not be physically meaningful, since the WCO$_2$  and SCO$_2$ spectrum should be strongly correlated due to the C-O orbital hybridization in the CO$_2$ spectral. Between the O$_2$ spectrum and CO$_2$ spectrum, there is no such physical justification for their strong cross-correlation. Therefore, we implement the Separable Emulator for the WCO$_2$ band the SCO$_2$ band. Based on the same settings in the Independent Emulator with the NNGP, we show the predictive summaries in Table~\ref{table: predictive measures (AS) IND vs SEP}. We found that the Independent Emulator gives slightly better predictive performance in terms of RMSPE, P$_{CI}(95\%)$, and CRPS than the Separable Emulator for the WCO$_2$, while it is the other way around for the SCO$_2$. Both of them gives satisfactory results. It appears that there is no obvious improvements in terms of these predictive summaries by taking into account the cross-correlation between the WCO$_2$ spectrum and SCO$_2$ spectrum. 

\begin{table}[hbt!]
\centering
\normalsize
   \caption{Numerical comparison of held-out sets based on the independent emulator and the separable emulator for WCO$_2$ and SCO$_2$ bands with the active subspace approach. The predictions are obtained at $n^*=1,000$ new input values. For each spectrum band, we compared RMSPE, $P_{CI}{(95\%)}$, and CRPS. The results are reported for both functional principal scores and radiance data.}
  {\resizebox{1.0\textwidth}{!}{%
  \setlength{\tabcolsep}{1.8em}
   \begin{tabular}{l c c c c c c} 
   \toprule %\noalign{\vskip .5pt}
& \multicolumn{3}{c}{Independent Emulator}  & \multicolumn{3}{c}{Separable Emulator} \\ \noalign{\vskip .5pt} \cmidrule(lr){2-4} \cmidrule(lr){5-7} %\noalign{\vskip 2.5pt} 
 & RMSPE & $P_{CI}{(95\%)}$  & CRPS & RMSPE & $P_{CI}{(95\%)}$  & CRPS \\ \noalign{\vskip .5pt} \hline %\noalign{\vskip 2.5pt} 
 \multicolumn{7}{c}{Results for each functional principal score} \\  \noalign{\vskip .5pt}   \hline \noalign{\vskip .5pt}
WCO$_2$ (PC 1) & 0.0341 &   0.952  &  0.0180 & 0.0345 & 0.950 & 0.0183 \\ \noalign{\vskip .5pt} 
SCO$_2$ (PC 1) & 0.0700  &  0.945  &  0.0375 & 0.0695 & 0.942 & 0.0373 \\ \noalign{\vskip .5pt} \hline %\noalign{\vskip 2.5pt}
\multicolumn{7}{c}{Results for radiances at each spectrum band} \\ \noalign{\vskip .5pt} \hline %\noalign{\vskip 2.5pt}
WCO$_2$  & 0.2429 & 0.950 & 0.2095 &0.2464 &0.949 &0.2140 \\ \noalign{\vskip .5pt} 
SCO$_2$  & 0.4103 & 0.932 & 0.3780 &0.4076  &0.934 &0.3573 \\
%\noalign{\vskip 1.5pt} 
\bottomrule
   \end{tabular}%
   }}
   \label{table: predictive measures (AS) IND vs SEP}
\end{table}

\subsection{Model Validation with a Reduced Order Model} \label{sec: validate with ROM}

Surrogate models have been developed to approximate expensive computer models from a statistical perspective and mathematical perspective in the UQ community. The emulator developed in this article is based on a statistical model that allows assessment of uncertainties in predicting the real-world physical process in the OCO-2 application, while the ROM introduced by \cite{HobbsBraverman2017} is formulated by simplifying the physical laws in the real-world process. It is of fundamental and practical interest to make comparison between these two types of surrogate models. In this section, we only focus on the comparison of predictive performance between the statistical emulators and the ROM. 

\begin{figure}[hbt!]
    \captionsetup[figure]{justification=centering}
\makebox[\textwidth][c]{ \includegraphics[width=1.0\linewidth, height=0.5\textheight]{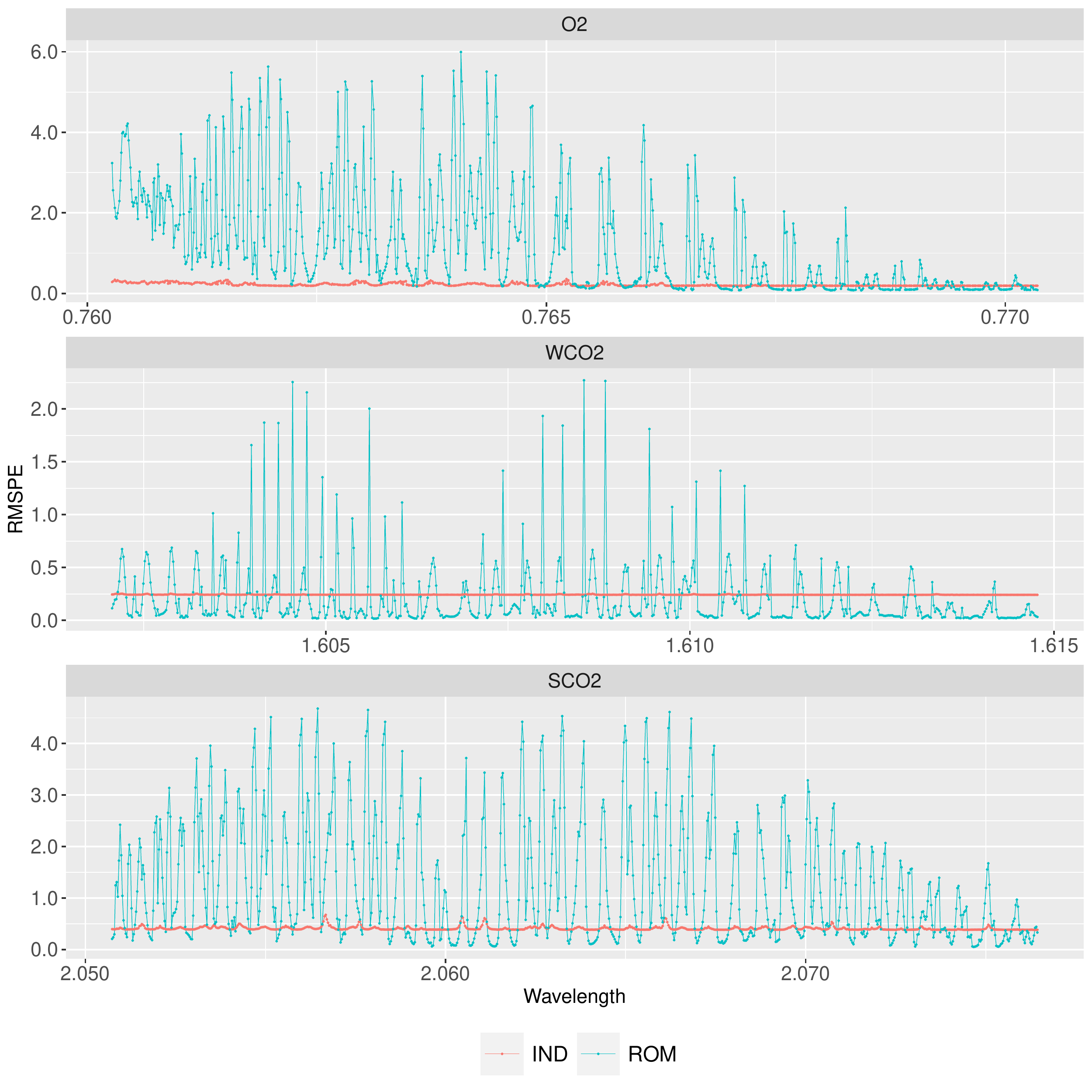}}
\caption{Comparison of pointwise RMSPE for predicted radiances based on the independent emulator (IND) and the reduced order model (ROM). The x-axis represents wavelength and the y-axis represents the RMSPE averaged over $n^*=1,000$ testing inputs.}
\label{fig: pointwise RMSPE}
\end{figure}

Following Section~\ref{sec: comparison with PCA and LaGP}, we run the ROM over these 1,000 held-out inputs. The evaluation of the ROM for a single sounding takes approximately 2 seconds on a single 2.6 GHz processor that is part of the OCO-2 operational computing cluster, while the evaluation of prediction based on the proposed emulators takes about 0.01 seconds on a 2013 Macbook Pro with 2.6 GHz processor for a single sounding once the model parameters are assumed to be fixed and known in MATLAB. With 3000 posterior samples of model parameters, the total computing time to make prediction for one sounding based on the proposed emulators is about 3.9 seconds on the 2013 Macbook Pro. Although prediction from the proposed emulators based on the fully Bayesian approach is more expensive, it can take into account the uncertainty in correlation parameters and nugget parameters. The computing time for prediction based on LaGP is similar to the proposed emulators based on NNGP, since they use the same number of nearest neighbors. For convenience, we refer to the Independent Emulator as IND and refer to the Separable Emulator as SEP.
We mainly compare the RMSPE obtained based on IND and ROM for radiance data. Figure~\ref{fig: pointwise RMSPE} shows the pointwise RMSPE based on IND and ROM that are averaged over all the new inputs for each wavelength. We can see that the IND gives very similar RMSPE across all the wavelengths. This is partly because the training inputs cover the input entire space very well and testing inputs are randomly selected in the input space (i.e., that is no place where extrapolation is involved), and partly because the emulator can capture the dependence structure between input space and output space very well.  The comparison of overall RMSPE is given in Figure~\ref{fig: NNGP and ROM} of the Supplementary Material.

\begin{figure}[hbt!]
\makebox[\textwidth][c]{ \includegraphics[width=1.0\linewidth, height=0.8\textheight]{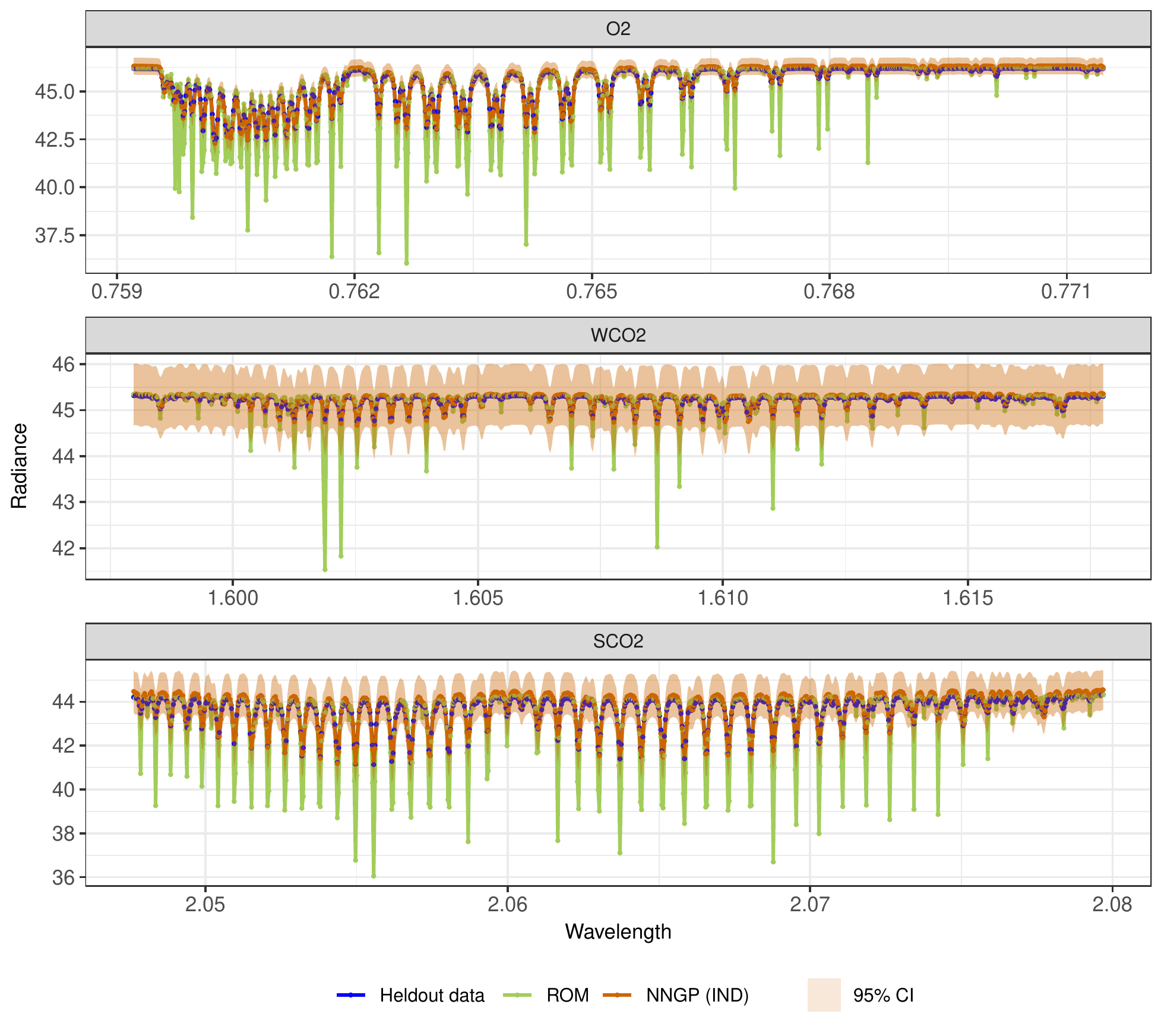}}
\caption{Comparison of the FP forward model simulated radiances, the ROM predictions, and the predicted radiances with their 95\% credible intervals from the independent emulator (IND).}
\label{fig: NNGP vs Surrogate}
\end{figure}

Figure~\ref{fig: NNGP vs Surrogate} shows that the predictive mean from the IND is closer to the true radiances from the FP forward model than those from the ROM. {The ROM produces noticeably lower radiances at certain wavelengths in all three bands. This discrepancy is likely due to the computational fidelity of the instrument model used in the ROM versus the FP forward model. The instrument model computations add a substantial computational cost to the FP forward model, and this component was simplified extensively in the ROM \citep{boesch2017, HobbsBraverman2017}}.  In terms of prediction uncertainty, the 95\% prediction intervals suggest that the IND can effectively quantify the prediction uncertainty across all the three bands. 
The prediction intervals in the O$_2$ band seem to be more narrower at a first glimpse than the prediction intervals in the WCO$_2$ and SCO$_2$ bands, however, a separate figure in Figure~\ref{fig: O2 prediction} of the Supplementary Material indicates that  the 95\% posterior prediction intervals can well capture the held-out data for the O$_2$ band. In addition,  the prediction performance based on the separable emulator in Figure~\ref{fig: SEP vs ROM} of the Supplementary Material also suggests that the separable emulator also give better prediction performance than the ROM.

\section{Discussion} \label{sec: conclusion}

This article focuses on an important scientific application in remote sensing science for uncertainty quantification of remote sensing retrievals. As current assessment of remote sensing retrievals relies on the full-physics forward model, uncertainty quantification for the retrieval problem via OSUEs can be computationally too demanding even with modern high-performance computing resources. To tackle this problem, a reduced order model has been developed to preserves key physical laws in the physical process of interest in the OCO-2 application; see \cite{HobbsBraverman2017}.  This article takes a statistical approach to constructing an emulator that can inherently assess the accuracy of the approximation to an expensive physical forward model with probabilistic uncertainties. This emulator not only allows fully Bayesian inference but also  provides a fast probabilistic approximation so that uncertainty quantification of remote sensing retrievals can be facilitated. This methodology can be readily extended to other remote sensing instruments.

The proposed emulator is formulated in three steps, resulting in a computationally efficient non-separable GP model with high-dimensional functional output. First, we use functional principal component analysis to deal with high-dimensional output at irregularly-spaced wavelengths with missing values. Second, we use the gradient information to reduce the input space dimension via the active subspace approach that induces a geometrically anisotropic correlation function to model each input dimension. Third, we adapt a computationally efficient NNGP in the computer model framework and compare it to the local GP approach \citep{Gramacy2015}. This formulation is applied for each spectral band, resulting in an independent emulator across different bands. To account for  the cross-correlation across spectral bands, we have formulated an upgraded version of the separable model based on a NNGP model that can deal with large-scale computer experiments. Our theoretical and empirical results suggest that such separable emulator give similar predictive performance as the independent emulator, although cross-covariances across different bands are accounted for. The current approach to approximating the functional output does not take into account the truncation error explicitly. This can be further improved by adding a white-noise error process. The dimension reduction approaches used in the output space and input space are currently performed separately, which allows fast computations. A simultaneous joint dimension reduction approach for both input space and output space can be appealing, but this can be challenging.  

The proposed emulator has demonstrated effectiveness in the OCO-2 application. Cross-validation studies have been used to compare the statistical emulator with other statistical methods and a reduced order model. Validating the statistical emulators with an existing physics-based surrogate model developed by \cite{HobbsBraverman2017} is particularly interesting in the UQ community. Our results suggest that the statistical emulator has very competitive performance and even gives better prediction results than the reduced order model. This is especially encouraging in the sense that a statistical emulator that is purely data-driven can be advantageous in remote sensing applications over a reduced order model that is built based on simplified physical laws in the real-world process.

\section*{Supplementary Material}
The Supplementary Material consists of additional technical descriptions and illustrative results.

\section*{Acknowledgements}
This material was based upon work partially supported by the National Science Foundation under Grant DMS-1638521 to the Statistical and Applied Mathematical Sciences Institute (SAMSI). Part of this work was performed at the Jet Propulsion Laboratory, California Institute of Technology, under contract with NASA. Any opinions, findings, and conclusions or recommendations expressed in this material are those of the author(s) and do not necessarily reflect the views of the National Science Foundation and NASA. This research began under the auspices and support of the SAMSI 2017-2018 research program on Mathematical and Statistical Methods for Climate and the Earth System (CLIM). The authors are grateful to Amy Braverman for introducing the OCO-2 application and sharing her comments and suggestions on this research.  The authors would also like to thank the Editor, the Associate Editor, and three anonymous referees for their constructive comments that improved the paper.

\section*{Appendix}
\appendix
\section{Correlation Functions} \label{app: correlation}
This section gives the forms of various correlation functions that can be used in Section~\ref{sec: Model}. 
Let $\bfs, \bfs'$ be any two inputs in a $p$-dimensional Euclidean space. The following correlation functions can be written as functions of a scaled distance between $\bfs$ and $\bfs'$:
\begin{itemize}
\item The Mat\'ern family \citep{Matern1960} has the form:
\begin{align*}
    \mathcal{M}(\bfs, \bfs'; \nu, \bftheta) = \frac{2^{1-\upsilon}}{\Gamma(\upsilon)} \left({\sqrt{2\upsilon}d} \right)^{\upsilon} \mathcal{K}_{\upsilon}\left({\sqrt{2\upsilon }d}\right),
\end{align*}
where $d=\sqrt{\sum_{i=1}^p (s_i-s_i')^2/\theta_i^2}$ is the distance between two inputs. $\mathcal{K}_{\upsilon}(\cdot)$ is the modified Bessel function of the second kind, $\theta_i$ is a range parameter, and $\upsilon$ is a smoothness parameter controlling the differentiability of process realizations. This correlation function is an anisotropic Mat\'ern correlation function.
\item The power exponential family has the form:
\begin{align*}
    \rho(\bfs, \bfs') = \exp\{ - d^{\alpha} \},
\end{align*}
where $d=\sqrt{\sum_{i=1}^p (s_i-s_i')^2/\theta_i^2}$ is the distance between two inputs. $\theta_i$ is a range parameter. $\alpha$ is the smoothness parameter. When $\alpha=2$, the correlation becomes an anisotropic Gaussian correlation function; when $\alpha=1$, the correlation becomes an anisotropic exponential correlation function.
\item The Confluent Hypergeometric family \citep{Ma2019cov} has the form:
\begin{align*}
\mathcal{CH}(\bfs, \bfs') = \frac{\Gamma(\nu+\alpha)}{\Gamma(\nu)} \mathcal{U}(\alpha, 1-\nu, d^2),
\end{align*}
where $d=\sqrt{\sum_{i=1}^p (s_i-s_i')^2/\theta_i^2}$ is the distance between two inputs. $\alpha$ is the tail decay parameter that controls the rate of polynomial decay. $\theta_i$ is a range parameter. $\nu$ is the smoothness parameter that controls the differentiability of process realizations. $\mathcal{U}$ is the confluent hypergeometric function of the second kind. This correlation function is an anisotropic Confluent Hypergeometric correlation function.

\end{itemize}

\section{Computational Details} \label{app: computation}
In Section~\ref{sec: Emulation}, both the independent emulator and the separable emulator require the posterior inference based on the following form of posterior distribution (with abused notations):
\begin{align*} %\label{eqn: post}
\pi(\bftheta \mid \bfZ) \propto \pi(\bftheta) |\bfR|^{-q/2} |\bfH^\top\bfR^{-1} \bfH|^{-q/2} |\bfZ^\top \bfG\bfZ|^{-(n-p)/2}, 
\end{align*}
where $\bfG:=\bfR^{-1} - \bfR^{-1} \bfH(\bfH^\top \bfR^{-1} \bfH)^{-1} \bfH^\top \bfR^{-1}$. For the independent emulator, $q=1$ and $\bfZ=\bfz_{k,i}$. For the separable emulator $\bfZ$ is a $n$-by-$q$ matrix of outputs. With the NNGP induced correlation function $c(\cdot, \cdot)$, we have
 \begin{align*}
 \bfR:=\bfC + \tau^2\bfI \approx \widetilde{\bfC} + \tau^2 \bfI,
\end{align*}
with $\widetilde{\bfC}:=(\bfI - \bfA)^{-1}\bfD (\bfI -\bfA)^{-\top}$ derived from the NNGP prior, where $\bfA$ is a lower triangular matrix and $\bfD$ is a diagonal matrix. Let $\bfOmega:= \widetilde{\bfC}^{-1}+ 1/\tau^2 \bfI$. Then  the inverse of $\bfR$ is $\bfR^{-1}=\bfOmega^{-1} (1/\tau^2) \widetilde{\bfC}^{-1}$. The determinant of $\bfR$ is $|\bfR|=|\bfOmega| |\bfD| (\tau^2)^n$. Notice that $ \widetilde{\bfC}^{-1}$ is a sparse matrix and so is $\bfOmega$. Thus $\bfH^\top \bfR^{-1}\bfH$ and $\bfZ^\top \bfG\bfZ$ can be computed efficiently by taking advantage of sparse matrix techniques. 

\section{Proof of Theorem~\ref{thm: irrelevance}} \label{app: theorem}
\begin{proof}
It follows from standard multivariate normal theory \citep[e.g.,][]{Conti2010} that under the separable covariance model, the posterior predictive distribution is 
\begin{align*}
\mathcal{M}_0: \bfz(\bfell_0) \mid \bftheta, \tau^2, \bfZ \sim \mathcal{MVT}(\hat{z}(\bfell_0), \hat{r}(\bfell_0) \hat{\bfSigma}_{gls}, n-p).
\end{align*}
This posterior predictive distribution is given in Equation~\eqref{eqn: prediction under SEP}.
For the independent emulator with the same input correlation parameters, the posterior predictive distribution is 
\begin{align*}
\mathcal{M}_1: \bfz(\bfell_0) \mid \bftheta, \tau^2, \bfZ \sim \mathcal{MVT}(\hat{z}(\bfell_0), \hat{r}(\bfell_0) \hat{\bfT}, n-p),
\end{align*}
where $\hat{\bfT} = \text{diag}( \hat{\bfSigma}_{gls})$. 
Clearly, the marginal posterior predictive distributions of $z_{k,i}(\Ell_0)$ given $\bftheta, \tau^2, \bfZ$ are the same under $\mathcal{M}_0$ and $\mathcal{M}_1$. 
\end{proof}

\bibliographystyle{apa}

%\singlespacing 
\begin{singlespace}
\setlength{\bibsep}{5pt}
\bibliography{emulator}

\begin{thebibliography}{}

\bibitem[\protect\astroncite{Bayarri et~al.}{2007}]{Bayarri2007}
Bayarri, M.~J., Berger, J.~O., Cafeo, J., Garcia-Donato, G., Liu, F., Palomo,
  J., Parthasarathy, R.~J., Paulo, R., Sacks, J., and Walsh, D. (2007).
\newblock {Computer model validation with functional output}.
\newblock {\em The Annals of Statistics}, 35(5):1874--1906.

\bibitem[\protect\astroncite{Berger et~al.}{2001}]{Berger2001}
Berger, J.~O., De~Oliveira, V., and Sanso, B. (2001).
\newblock {Objective Bayesian analysis of spatially correlated data}.
\newblock {\em Journal of the American Statistical Association},
  96(456):1361--1374.

\bibitem[\protect\astroncite{Bilionis et~al.}{2013}]{BILIONIS2013212}
Bilionis, I., Zabaras, N., Konomi, B.~A., and Lin, G. (2013).
\newblock Multi-output separable {Gaussian} process: Towards an efficient,
  fully {Bayesian} paradigm for uncertainty quantification.
\newblock {\em Journal of Computational Physics}, 241:212 -- 239.

\bibitem[\protect\astroncite{B\"{o}sch et~al.}{2017}]{boesch2017}
B\"{o}sch, H., Brown, L., Castano, R., Christi, M., Connor, B., Crisp, D.,
  Eldering, A., Fisher, B., Frankenberg, C., Gunson, M., Granat, R., et~al.
  (2017).
\newblock {Orbiting Carbon Observatory (OCO-2) Level 2 Full Physics Algorithm
  Theoretical Basis Document}.
\newblock URL:
  \url{https://docserver.gesdisc.eosdis.nasa.gov/public/project/OCO/OCO2_L2_ATBD.V8.pdf}.

\bibitem[\protect\astroncite{Braverman et~al.}{2019}]{Braverman2019}
Braverman, A., Hobbs, J., Teixeira, J., and Gunson, M. (2019).
\newblock Post hoc uncertainty quantification for remote sensing observing
  systems.
\newblock Submitted.

\bibitem[\protect\astroncite{Connor et~al.}{2016}]{ConnorEtAl2016}
Connor, B., B\"{o}sch, H., McDuffie, J., Taylor, T., Fu, D., Frankenberg, C.,
  O'Dell, C., Payne, V.~H., Gunson, M., Pollock, R., Hobbs, J., Oyafuso, F.,
  and Jiang, Y. (2016).
\newblock Quantification of uncertainties in {OCO}-2 measurements of {XCO}$_2$:
  simulations and linear error analysis.
\newblock {\em Atmospheric Measurement Techniques}, 9.

\bibitem[\protect\astroncite{Constantine and Gleich}{2014}]{Constantine2014MC}
Constantine, P. and Gleich, D. (2014).
\newblock Computing active subspaces with {Monte Carlo}.
\newblock {\em arXiv.org}, page arXiv:1408.0545.

\bibitem[\protect\astroncite{Constantine et~al.}{2014}]{Constantine2014}
Constantine, P.~G., Dow, E., and Wang, Q. (2014).
\newblock Active subspace methods in theory and practice: Applications to
  kriging surfaces.
\newblock {\em SIAM Journal on Scientific Computing}, 36(4):A1500--A1524.

\bibitem[\protect\astroncite{Constantine et~al.}{2017}]{Constantine2017}
Constantine, P.~G., Eftekhari, A., Hokanson, J., and Ward, R.~A. (2017).
\newblock A near-stationary subspace for ridge approximation.
\newblock {\em Computer Methods in Applied Mechanics and Engineering},
  326(Supplement C):402 -- 421.

\bibitem[\protect\astroncite{Constantine et~al.}{2016}]{Constantine2016}
Constantine, P.~G., Kent, C., and Bui-Thanh, T. (2016).
\newblock Accelerating {Markov Chain Monte Carlo} with active subspaces.
\newblock {\em SIAM Journal on Scientific Computing}.

\bibitem[\protect\astroncite{Conti et~al.}{2009}]{Conti2009}
Conti, S., Gosling, J.~P., Oakley, J.~E., and O'Hagan, A. (2009).
\newblock {Gaussian process emulation of dynamic computer codes}.
\newblock {\em Biometrika}, 96(3):663--676.

\bibitem[\protect\astroncite{Conti and O'Hagan}{2010}]{Conti2010}
Conti, S. and O'Hagan, A. (2010).
\newblock {Bayesian emulation of complex multi-output and dynamic computer
  models}.
\newblock {\em Journal of Statistical Planning and Inference}, 140(3):640--651.

\bibitem[\protect\astroncite{Cook}{1994}]{Cook1994}
Cook, R.~D. (1994).
\newblock On the interpretation of regression plots.
\newblock {\em Journal of the American Statistical Association}, 89(425):177.

\bibitem[\protect\astroncite{Cook}{2009}]{Cook2009}
Cook, R.~D. (2009).
\newblock {\em Regression Graphics: Ideas for Studying Regressions Through
  Graphics}, volume 482.
\newblock John Wiley \& Sons, New York, NY.

\bibitem[\protect\astroncite{Cressie}{1993}]{Cressie1993}
Cressie, N. (1993).
\newblock {\em {Statistics for Spatial Data}}.
\newblock John Wiley {\&} Sons, New York, revised edition.

\bibitem[\protect\astroncite{Cressie}{2018}]{cressiejasa}
Cressie, N. (2018).
\newblock {Mission {CO$_2$}ntrol: A statistical scientist's role in remote
  sensing of carbon dioxide}.
\newblock {\em Journal of the American Statistical Association}, 113:152--168.

\bibitem[\protect\astroncite{Cressie and Johannesson}{2008}]{Cressie2008}
Cressie, N. and Johannesson, G. (2008).
\newblock {Fixed rank kriging for very large spatial data sets}.
\newblock {\em Journal of the Royal Statistical Society: Series B (Statistical
  Methodology)}, 70(1):209--226.

\bibitem[\protect\astroncite{Datta et~al.}{2016}]{Datta2016}
Datta, A., Banerjee, S., Finley, A.~O., and Gelfand, A.~E. (2016).
\newblock Hierarchical nearest-neighbor {Gaussian} process models for large
  geostatistical datasets.
\newblock {\em Journal of the American Statistical Association},
  111(514):800--812.

\bibitem[\protect\astroncite{Eldering et~al.}{2017}]{ElderingAMT2017}
Eldering, A., O'Dell, C.~W., Wennberg, P.~O., Crisp, D., Gunson, M., Viatte,
  C., Avis, C., Yoshimizu, J., et~al. (2017).
\newblock {The Orbiting Carbon Observatory-2: First 18 months of science data
  products}.
\newblock {\em Atmospheric Measurement Techniques}, 10(2):549--563.

\bibitem[\protect\astroncite{Finley et~al.}{2019}]{Finley2019}
Finley, A.~O., Datta, A., Cook, B.~D., Morton, D.~C., Andersen, H.~E., and
  Banerjee, S. (2019).
\newblock Efficient algorithms for {B}ayesian nearest neighbor {G}aussian
  processes.
\newblock {\em Journal of Computational and Graphical Statistics},
  28(2):401--414.

\bibitem[\protect\astroncite{Fricker et~al.}{2013}]{Fricker2013}
Fricker, T.~E., Oakley, J.~E., and Urban, N.~M. (2013).
\newblock Multivariate {Gaussian} process emulators with nonseparable
  covariance structures.
\newblock {\em Technometrics}, 55(1):47--56.

\bibitem[\protect\astroncite{Gneiting and Raftery}{2007}]{Gneiting2007}
Gneiting, T. and Raftery, A.~E. (2007).
\newblock Strictly proper scoring rules, prediction, and estimation.
\newblock {\em Journal of the American Statistical Association},
  102(477):359--378.

\bibitem[\protect\astroncite{Gramacy}{2016}]{Gramacy2016}
Gramacy, R. (2016).
\newblock {laGP}: Large-scale spatial modeling via local approximate {Gaussian}
  processes in \textsf{R}.
\newblock {\em Journal of Statistical Software, Articles}, 72(1):1--46.

\bibitem[\protect\astroncite{Gramacy and Apley}{2015}]{Gramacy2015}
Gramacy, R.~B. and Apley, D.~W. (2015).
\newblock Local {Gaussian} process approximation for large computer
  experiments.
\newblock {\em Journal of Computational and Graphical Statistics},
  24(2):561--578.

\bibitem[\protect\astroncite{Gu and Berger}{2016}]{Gu2016}
Gu, M. and Berger, J.~O. (2016).
\newblock {Parallel partial Gaussian process emulation for computer models with
  massive output}.
\newblock {\em The Annals of Applied Statistics}, 10(3):1317--1347.

\bibitem[\protect\astroncite{Guillas et~al.}{2018}]{Guillas2018}
Guillas, S., Sarri, A., Day, S.~J., Liu, X., and Dias, F. (2018).
\newblock {Functional emulation of high resolution tsunami modelling over
  Cascadia}.
\newblock {\em The Annals of Applied Statistics}, 12(4):2023--2053.

\bibitem[\protect\astroncite{Gurney et~al.}{2002}]{Gurney2002}
Gurney, K.~R., Law, R.~M., Denning, A.~S., Rayner, P.~J., Baker, D., Bousquet,
  P., Bruhwiler, L., Chen, Y.~H., Ciais, P., Fan, S., Fung, I.~Y., Gloor, M.,
  Heimann, M., Higuchi, K., John, J., Maki, T., Maksyutov, S., Masarie, K.,
  Peylin, P., Prather, M., Pak, B.~C., Randerson, J., Sarmiento, J., Taguchi,
  S., Takahashi, T., and Yuen, C.-W. (2002).
\newblock {Towards robust regional estimates of CO2 sources and sinks using
  atmospheric transport models}.
\newblock {\em Nature}, 415(6872):626--630.

\bibitem[\protect\astroncite{Higdon}{2002}]{Higdon2002}
Higdon, D. (2002).
\newblock Space and space-time modeling using process convolutions.
\newblock In Anderson, C.~W., Barnett, V., Chatwin, P.~C., and El-Shaarawi,
  A.~H., editors, {\em Quantitative Methods for Current Environmental Issues},
  pages 37--56, London. Springer London.

\bibitem[\protect\astroncite{Higdon et~al.}{2008}]{Higdon2008}
Higdon, D., Gattiker, J., and Williams, B. (2008).
\newblock {Computer model calibration using high-dimensional output}.
\newblock {\em Journal of the American Statistical Association},
  103(482):570--583.

\bibitem[\protect\astroncite{Hobbs et~al.}{2017}]{HobbsBraverman2017}
Hobbs, J., Braverman, A., Cressie, N., Granat, R., and Gunson, M. (2017).
\newblock Simulation-based uncertainty quantification for estimating
  atmospheric {CO}$_2$ from satellite data.
\newblock {\em SIAM/ASA Journal on Uncertainty Quantification}, 5(1):956--985.

\bibitem[\protect\astroncite{Journel and Huijbregts}{1978}]{Journel1978}
Journel, A.~G. and Huijbregts, C.~J. (1978).
\newblock {\em Mining Geostatistics}.
\newblock Academic Press.

\bibitem[\protect\astroncite{Karhunen}{1947}]{karhunen1947lineare}
Karhunen, K. (1947).
\newblock {\em {\"U}ber lineare Methoden in der Wahrscheinlichkeitsrechnung},
  volume~37.
\newblock Sana.

\bibitem[\protect\astroncite{Katzfuss and Guinness}{2020}]{Katzfuss2020}
Katzfuss, M. and Guinness, J. (2020).
\newblock A general framework for {V}ecchia approximations of {G}aussian
  processes.
\newblock {\em Statistical Science}.
\newblock To appear.

\bibitem[\protect\astroncite{Kennedy and O'Hagan}{2001}]{Kennedy2001}
Kennedy, M.~C. and O'Hagan, A. (2001).
\newblock {Bayesian calibration of computer models}.
\newblock {\em Journal of the Royal Statistical Society: Series B (Statistical
  Methodology)}, 63(3):425--464.

\bibitem[\protect\astroncite{Li}{1991}]{Li1991}
Li, K.-C. (1991).
\newblock Sliced inverse regression for dimension reduction.
\newblock {\em Journal of the American Statistical Association}, 86(414):316.

\bibitem[\protect\astroncite{Liu and Guillas}{2017}]{Liu2017}
Liu, X. and Guillas, S. (2017).
\newblock Dimension reduction for {Gaussian} process emulation: An application
  to the influence of bathymetry on tsunami heights.
\newblock {\em SIAM/ASA Journal on Uncertainty Quantification}, 5(1):787--812.

\bibitem[\protect\astroncite{Ma and Bhadra}{2019}]{Ma2019cov}
Ma, P. and Bhadra, A. (2019).
\newblock Kriging: {B}eyond {M}at\'ern.
\newblock arXiv preprint arXiv:1911.05865.
\newblock URL: https://arxiv.org/abs/1911.05865.

\bibitem[\protect\astroncite{Mak et~al.}{2018}]{Mak2018}
Mak, S., Sung, C.-L., Wang, X., Yeh, S.-T., Chang, Y.-H., Joseph, V.~R., Yang,
  V., and Wu, C. F.~J. (2018).
\newblock An efficient surrogate model for emulation and physics extraction of
  large eddy simulations.
\newblock {\em Journal of the American Statistical Association},
  113(524):1443--1456.

\bibitem[\protect\astroncite{Mat{\'e}rn}{1960}]{Matern1960}
Mat{\'e}rn, B. (1960).
\newblock \emph{Spatial variation}, {Meddelanden fran Statens
  Skogsforskningsinstitut, 49, 5. Second ed. (1986), Lecture Notes in
  Statistics 36, New York: Springer}.

\bibitem[\protect\astroncite{Patra et~al.}{2017}]{Patra2017}
Patra, P.~K., Crisp, D., Kaiser, J.~W., Wunch, D., Saeki, T., Ichii, K.,
  Sekiya, T., Wennberg, P.~O., Feist, D.~G., Pollard, D.~F., Griffith, D.
  W.~T., Velazco, V.~A., De~Maziere, M., Sha, M.~K., Roehl, C., Chatterjee, A.,
  and Ishijima, K. (2017).
\newblock {The Orbiting Carbon Observatory (OCO-2) tracks 2{\textendash}3
  peta-gram increase in carbon release to the atmosphere during the
  2014{\textendash}2016 El Ni{\~n}o}.
\newblock {\em Scientific Reports}, 7(1):13567.

\bibitem[\protect\astroncite{Ramsay and Silverman}{2005}]{ramsaysilverman2005}
Ramsay, J. and Silverman, B.~W. (2005).
\newblock {\em Functional Data Analysis}.
\newblock Springer, New York, NY.

\bibitem[\protect\astroncite{Ramsay et~al.}{2018}]{fda}
Ramsay, J.~O., Wickham, H., Graves, S., and Hooker, G. (2018).
\newblock {\em {fda: Functional Data Analysis}}.
\newblock R package version 2.4.8.

\bibitem[\protect\astroncite{Rodgers}{2000}]{Rodgers2000}
Rodgers, C.~D. (2000).
\newblock {\em Inverse Methods for Atmospheric Sounding}.
\newblock World Scientific.

\bibitem[\protect\astroncite{Sacks et~al.}{1989}]{Sacks1989}
Sacks, J., Welch, W.~J., Mitchell, T.~J., and Wynn, H.~P. (1989).
\newblock Design and analysis of computer experiments.
\newblock {\em Statistical Science}, 4(4):409--423.

\bibitem[\protect\astroncite{Santner et~al.}{2018}]{Santner2018}
Santner, T.~J., Williams, B.~J., and Notz, W.~I. (2018).
\newblock {\em {The design and analysis of computer experiments; 2nd ed.}}
\newblock Springer series in statistics. Springer, New York, NY.

\bibitem[\protect\astroncite{Stein}{2014}]{Stein2014}
Stein, M.~L. (2014).
\newblock {Limitations on low rank approximations for covariance matrices of
  spatial data}.
\newblock {\em Spatial Statistics}, 8:1--19.

\bibitem[\protect\astroncite{Turmon and Braverman}{2019}]{Turmon2019}
Turmon, M. and Braverman, A. (2019).
\newblock Uncertainty quantification for {JPL} retrievals.
\newblock Technical report, Pasadena, CA: Jet Propulsion Laboratory, National
  Aeronautics and Space Administration, 2019.
\newblock http://hdl.handle.net/2014/45978.

\bibitem[\protect\astroncite{Vecchia}{1988}]{vecchia1988estimation}
Vecchia, A.~V. (1988).
\newblock Estimation and model identification for continuous spatial processes.
\newblock {\em Journal of the Royal Statistical Society. Series B},
  66(2):297--312.

\bibitem[\protect\astroncite{Welch et~al.}{1992}]{Welch1992}
Welch, W.~J., Buck, R.~J., Sacks, J., Wynn, H.~P., Mitchell, T.~J., and Morris,
  M.~D. (1992).
\newblock Screening, predicting, and computer experiments.
\newblock {\em Technometrics}, 34(1):15.

\bibitem[\protect\astroncite{Wunch et~al.}{2011}]{Wunch2011}
Wunch, D., Toon, G.~C., Blavier, J.-F.~L., Washenfelder, R.~A., Notholt, J.,
  Connor, B.~J., Griffith, D. W.~T., Sherlock, V., and Wennberg, P.~O. (2011).
\newblock {The Total Carbon Column Observing Network}.
\newblock {\em Philosophical Transactions of the Royal Society A: Mathematical,
  Physical and Engineering Sciences}, 369(1943):2087--2112.

\bibitem[\protect\astroncite{Zhang et~al.}{2018}]{zhangtechnom}
Zhang, B., Cressie, N., and Wunch, D. (2018).
\newblock Inference for errors-in-variables models in the presence of spatial
  and temporal dependence with an application to a satellite remote sensing
  campaign.
\newblock {\em Technometrics}, 61:187--201.

\bibitem[\protect\astroncite{Zimmerman}{1993}]{Zimmerman1993}
Zimmerman, D.~L. (1993).
\newblock {Another look at anisotropy in geostatistics}.
\newblock {\em Mathematical Geology}, 25(4):453--470.

\end{thebibliography}


\begin{thebibliography}{}

\bibitem[\protect\astroncite{Cressie and Johannesson}{2008}]{Cressie2008}
Cressie, N. and Johannesson, G. (2008).
\newblock {Fixed rank kriging for very large spatial data sets}.
\newblock {\em Journal of the Royal Statistical Society: Series B (Statistical
  Methodology)}, 70(1):209--226.

\bibitem[\protect\astroncite{Datta et~al.}{2016}]{Datta2016}
Datta, A., Banerjee, S., Finley, A.~O., and Gelfand, A.~E. (2016).
\newblock Hierarchical nearest-neighbor {Gaussian} process models for large
  geostatistical datasets.
\newblock {\em Journal of the American Statistical Association},
  111(514):800--812.

\bibitem[\protect\astroncite{Finley et~al.}{2019}]{Finley2019}
Finley, A.~O., Datta, A., Cook, B.~D., Morton, D.~C., Andersen, H.~E., and
  Banerjee, S. (2019).
\newblock Efficient algorithms for {B}ayesian nearest neighbor {G}aussian
  processes.
\newblock {\em Journal of Computational and Graphical Statistics},
  28(2):401--414.

\bibitem[\protect\astroncite{Karhunen}{1947}]{karhunen1947lineare}
Karhunen, K. (1947).
\newblock {\em {\"U}ber lineare Methoden in der Wahrscheinlichkeitsrechnung},
  volume~37.
\newblock Sana.

\bibitem[\protect\astroncite{Ramsay and Silverman}{2005}]{ramsaysilverman2005}
Ramsay, J. and Silverman, B.~W. (2005).
\newblock {\em Functional Data Analysis}.
\newblock Springer, New York, NY.

\bibitem[\protect\astroncite{Stein}{2014}]{Stein2014}
Stein, M.~L. (2014).
\newblock {Limitations on low rank approximations for covariance matrices of
  spatial data}.
\newblock {\em Spatial Statistics}, 8:1--19.

\bibitem[\protect\astroncite{Vecchia}{1988}]{vecchia1988estimation}
Vecchia, A.~V. (1988).
\newblock Estimation and model identification for continuous spatial processes.
\newblock {\em Journal of the Royal Statistical Society. Series B},
  66(2):297--312.

\end{thebibliography}
\end{singlespace}

%%%%%%%%%%%%%%%%%%%%%%%%%%%%%%%%%%%%%%%%%%%%%%%%%%%%%%%%%%%%%%
%%%%%%%%%%%%%%%%%%%%%%%%%%%%%%%%%%%%%%%%%%%%%%%%%%%%%%%%%%%%%%

\newpage
\appendix

%%%% Main text entry area:
\setcounter{equation}{1}
\setcounter{page}{1}
\setcounter{table}{0}
\setcounter{section}{0}
\setcounter{figure}{0}
%\numberwithin{table}{section}
\renewcommand{\theequation}{S.\arabic{section}.\arabic{equation}}
\renewcommand{\thesection}{S.\arabic{section}}
\renewcommand{\thesubsection}{S.\arabic{section}.\arabic{subsection}}
\renewcommand{\thepage}{S.\arabic{page}}
\renewcommand{\thetable}{S.\arabic{table}}
\renewcommand{\thefigure}{S.\arabic{figure}}
%\baselineskip=15pt

%\clearpage\pagebreak\newpage

\renewcommand{\theequation}{S.\arabic{section}.\arabic{equation}}
\renewcommand{\thesubsection}{S.\arabic{section}.\arabic{subsection}}

\begin{center}
	{\Large \textbf{Supplementary Material}} 
\end{center}
%\newpage 

The Supplementary Material consists of four sections. In Section~\ref{app: fpca}, we give the detailed description of the FPCA method. In Section~\ref{app: NNGP}, we give the detailed formulation for the nearest neighbor Gaussian process.  Section~\ref{app: PCA} includes results from the PCA approach for input space dimension reduction. Section~\ref{app: numeric} includes additional numerical results.

\section{Functional principal component analysis} \label{app: fpca}

The FP forward model output consists of radiances at 3,048 irregularly-spaced wavelengths. To respect the nature of spectral outputs, a functional data analysis approach is suitable to account for the correlations between radiances across wavelengths and to deal with irregularly sampled data with missing values \citep{ramsaysilverman2005}. The radiance vectors are transformed to a functional form (as a function of wavelength) by fitting cubic B-spline basis to the observed data. Then functional principal component analysis is applied to this functional data for dimension reduction.

The central idea of FPCA is to find the set of orthogonal functions, the so-called functional principal components, while retaining as much of the variation in the data as possible. Let $y(\bfx, \bfb; \omega)$ be a random function of $\omega$ on a compact interval $\mathcal{I}$ and belonging to the $L^{2}$ space. Mercer's theorem implies a spectral decomposition of the covariance function $V(\omega,\omega')$ for $y(\bfx, \bfb; \omega)$ and $y(\bfx, \bfb; \omega')$,
\begin{eqnarray}
V(\omega,\omega')=\sum_{l}^{\infty}\lambda_{l}\eta_{l}(\omega)\eta_{l}(\omega'),
\label{eq:K1}
\end{eqnarray}
where $\lambda_{1}\geq \lambda_{2} \geq \cdots \geq 0$ are the ordered eigenvalues and the $\eta_{k}(\cdot)$'s are the corresponding eigenfunctions. The Karhunen-Loeve expansion \citep{karhunen1947lineare} allows the representation of a random curve $y(\bfx, \bfb; \omega)$ as an infinite linear combination of orthogonal functions,
\begin{eqnarray}
y(\bfx, \bfb; \omega)=\sum_{l=1}^{\infty}z_{l}(\bfx, \bfb)\eta_{l}(\omega), 
\label{eq:KL1}
\end{eqnarray}
where the coefficient $z_{l}(\bfx,\bfb)=\int y(\bfx, \bfb; \omega)\eta_{l}(\omega)d\omega$ is called the the functional principal component score corresponding to the $l$th FPC $\eta_{l}(\omega)$. Note that $z_{l}$ are uncorrelated random variables with $E(z_{l})=0$ and $Var(z_{l})=\lambda_{l}$. The eigenvalues $\lambda_{l}$ measure the variation in $y(\bfx, \bfb; \omega)$ in the $\eta_{l}$ direction. FPCA can achieve dimension reduction by retaining only the first $p_y$ eigencomponents, eigenfunctions $\eta(\omega)$ and eigenvalues $\lambda$ in Equation~\eqref{eq:KL1}. 

Such a basis expansion has also been used in spatial statistics. One important distinction between functional data analysis and spatial data analysis is that in spatial statistics, there is no replication at each spatial location, that is, the number of trajectories or images observed is just one, while in functional data analysis, it is assumed that we have observed $n$ ($n>1$) trajectories. This is true in the application, since we have the input space that provides more than 10,000 trajectories. Therefore, such basis expansion does have any issues as the low-rank models \citep{Cressie2008} in spatial statistics as pointed out by \cite{Stein2014}.

To estimate the eigencomponents, mean and covariance functions for the functional data are first estimated. 
The estimated mean function is $\bar{y}(\omega)=1/n\sum_{j=1}^{n}y(\bfx_{j}, \bfb_{j}; \omega)$ and the estimator of the covariance function $V(\omega,\omega')$ is
%\begin{eqnarray}
$\hat{V}(\omega,\omega')=\frac{1}{n}\sum_{j=1}^{n}(y(\bfx_{j}, \bfb_{j}; \omega)-\bar{y}(\omega))(y(\bfx_{j}, \bfb_{j}; \omega')-\bar{y}(\omega'))$.
%\label{eq:Cov1}
%\end{eqnarray}
An empirical version of Equation~\eqref{eq:K1} is given by   
\begin{eqnarray}
\hat{V}(\omega,\omega')=\sum_{l}^{\infty}\hat{\lambda}_{l}\hat{\eta}_{l}(\omega)\hat{\eta}_{l}(\omega'),
\label{eq:EK1}
\end{eqnarray}
where $\hat{\lambda}_{1}\geq \hat{\lambda}_{2} \geq \cdots \geq 0$ and $\hat{\eta}_{k}(\cdot)$'s are the ordered eigenvalues and the corresponding eigenfunctions of $\hat{V}(\omega,\omega')$. These eigenvalues and eigenfunctions can be computed by solving the eigen equations 
\begin{equation}
  \int \hat{V}(\omega,\omega')\eta_l(\omega') d\omega' = \lambda_l \eta_l(\omega), l = 1,2, \ldots,
  \label{eq:eigeneq1}
\end{equation}

To implement the FPCA, the centered radiances, $y^{c}(\bfx_{j}, \bfb_{j}; \omega)=y(\bfx_{j}, \bfb_{j}; \omega) - \bar{y}(\omega)$, are transformed to a functional form using a basis function representation given by 
%\begin{equation}
    $y^{c}(\bfx_{j}, \bfb_{j}; \omega) = \sum_{g=1}^{G}a_{jg}\phi_g(\omega), j = 1,2, \ldots, n,$
%\end{equation}
where $\phi$'s are a series of basis functions and the $a_{jg}$'s are the corresponding coefficients. This expression can be written in  matrix notation, 
\begin{equation}
\mathbf{y}^c(\omega)=\mathbf{A}\boldsymbol{\phi}(\omega).
\label{Eq:basis11}
\end{equation}
There are many choices for $\boldsymbol{\phi}(\omega)$, such as polynomial basis, exponential basis, spline basis, Fourier basis etc. For the radiance data B-splines basis functions are used where the knot points are selected equidistantly over the range of wavelengths in each of the three bands. The coefficients $a_{jg}$'s are obtained using the method of least squares. A similar basis representation is also obtained for the mean radiance function, given by $\bar{y}(\omega)=\frac{1}{n}\sum_{g=1}^{G}\bar{a}_{g}\phi_g(\omega)$. Thus the mean function defined in Equation~\eqref{eqn: model for y} is estimated by $\hat{\mu}(\omega)=\frac{1}{n}\sum_{g=1}^{G}\bar{a}_{g}\phi_g(\omega)$, where $\bar{a}_{g}$'s are obtained using method of penalized least squares.
Substituting the basis representation in Equation~\eqref{Eq:basis11} in the covariance function representation in Equation~\eqref{eq:EK1} we have
\begin{equation}
\hat{V}(\omega,\omega')= n^{-1}\boldsymbol{\phi}^T(\omega)\mathbf{A}^T\mathbf{A}\boldsymbol{\phi}(\omega').
\end{equation}
Let the $l^{th}$ eigenfunction in Equation~\eqref{eq:EK1} be expressed as 
\begin{equation}
\hat{\eta}_l(\omega) = \mathbf{d}_l^T\boldsymbol{\phi}(\omega).
\label{eq:eta11}
\end{equation}
Substituting this in Equation~\eqref{eq:eigeneq1}, the eigenequations become 
\begin{equation}
     n^{-1}\boldsymbol{\phi}^T(\omega)\mathbf{A}^T\mathbf{A}\mathbf{J}\mathbf{d}_l = \lambda_l\boldsymbol{\phi}^T(\omega)\mathbf{d}_l, l=1,2,\ldots,  
     \label{eq:eigen21}
\end{equation}
where $\mathbf{J}= \int \boldsymbol{\phi}(\omega')\boldsymbol{\phi}^T(\omega') d\omega' $. Furthermore, Equation~\eqref{eq:eigen21}  has to be true for all argument values $\omega$, and consequently 
\begin{equation}
     n^{-1}\mathbf{A}^T\mathbf{A}\mathbf{J}\mathbf{d}_l = \lambda_l\mathbf{d}_l, l=1,2, \ldots, \ %\text{subject to the constraint} \ \mathbf{d}^T \mathbf{J} \mathbf{d}=1 
     \label{eq:eigen31}
\end{equation}
{subject to the constraint \ $\mathbf{d}_l^T \mathbf{J} \mathbf{d}_{l'}=\delta_{l,l'}$, where $\delta_{l,l'} = 1$ if $l=l'$ and $0$ otherwise.}
Equation~\eqref{eq:eigen31} is solved numerically to obtain the $\mathbf{d}_l$'s, which are then substituted in Equation~\eqref{eq:eta11} to obtain the eigenfunctions $\hat{\eta_l}(\omega)$. 
The functional principal component scores are then estimated by numerical integration of $\hat{z}_l(\bfx, \bfb)=\int y^c(\bfx, \bfb; \omega)\hat{\eta_l}(\omega)d\omega$.

\subsection{B-Spline Basis Functions}\label{app: B-Spline}

While implementing FPCA the radiance vectors are transformed to a functional form (as a function of wavelength) by fitting cubic B-spline basis to the observed data, i.e., in Equation~\eqref{Eq:basis11} we consider
%\begin{equation}
    $\phi_j(\omega)=N_{j,3}(\omega)$,
%\end{equation}
where, $N_{j,3}(\omega)$'s are obtained by the recursive formula 
\begin{eqnarray*}
    &N_{j,0}(\omega)=
    \begin{cases} 
    1, & \text{if} \ \omega_j < t<\omega_{j+1} \\
    0, & \text{otherwise}
    \end{cases} \\\\
    &N_{j,p}(\omega) = \frac{\omega-\omega_j}{\omega_{j+p}-\omega_j}N_{j,p-1} + \frac{\omega_{j+p+1}-\omega}{\omega_{j+p+1}-\omega_{j+1}}N_{j+1,p-1}, p =1,2,3.
\end{eqnarray*}
Here $\omega_1, \omega_2,...,\omega_{K+1}$ are the called the knot points and are selected equidistantly over the range of wavelengths in each of the three bands. A similar basis representation is also obtained for the mean radiance vector $\bar{y}(\omega)$.

\section{Nearest Neighbor Gaussian Process} \label{app: NNGP}

For nontational convenience, the subscripts are suppressed for the FPCA weights. Suppose that the FPCA weight $z(\bfell)$ on $\mathcal{L}$ is modeled with a Gaussian process with three different components:
\begin{eqnarray} \label{eqn: uNNGP}
  z(\bfell) =\bfh^\top(\bfell)\bfbeta + w(\bfell) + \epsilon(\bfell),\quad \bfell \in \mathcal{L},
\end{eqnarray}
where $\bfh(\cdot)$ is a vector of fixed basis function and $\bfbeta$ is the corresponding unknown regression coefficients.  $w(\cdot)$ is a Gaussian process with zero mean and covariance function $\sigma^2c(\cdot, \cdot)$ with correlation parameters $\bftheta$ and marginal variance parameter $\sigma^2$. $\epsilon(\cdot)$ is a Gaussian white-noise process accounting for the nugget effect with variance $\tau^2/\sigma^2$. This error term can potentially capture the truncation error in the basis function representation in Equation~\eqref{eqn: model for y}. The correlation function $c(\bfell, \bfell)$ is assumed to be a range anisotropic correlation function, see Appendix~\ref{app: correlation} for various examples. 

Given $z(\cdot)$ over  $n$ input locations $\mathcal{L}^O=\{\bfell_1, \dots , \bfell_n\} \subset \mathcal{L} $. Let $\bfz = (z(\bfell_1),z(\bfell_2), \dots ,z(\bfell _n))^{\top}$ be a vector of corresponding outputs. Then the likelihood function of $\bfz$ given model parameters $\bfbeta, \bftheta, \tau^2, \sigma^2$ is 
\begin{eqnarray} \label{eqn: likelihood}
L(\bfz \mid \bfbeta, \bftheta, \tau^2, \sigma^2) = \mathcal{N}_n( \bfH \bfbeta, \sigma^2 (\bfC+\tau^2\mathbf{I})) 
\end{eqnarray}
where $\bfH\equiv (\bfh(\bfell_1), \ldots, \bfh(\bfell_n))^\top$ and $\bfC = [c(\bfell_i, \bfell_j)]_{i,j=1, \ldots, n}$. 

Model fitting and prediction in the full GP model through either likelihood-based approaches typically require the repeated evaluation of the likelihood function, which involves calculating the inverse and determinant of large, dense, and unstructured $n$-by-$n$ covariance matrix with $O(n^3)$ flops and $O(n^2)$ storage. However, memory limitations and computational complexity grow with $n$ as well as with the dimensionality of the input space, making the full GP modeling impractical for large datasets.

To reduce the computational complexity of the full Gaussian process, many methods have been proposed to tackle this issue. We will employ the nearest-neighbor Gaussian process \citep[NNGP;][]{Datta2016} in this article. Specifically, the $\mNNGP$ extends Vecchia's approximation \citep{vecchia1988estimation} ideas based on localized information. To approximate $w(\cdot)$, the NNGP model is constructed in two steps. First, we specify a multivariate Gaussian distribution over a fixed set of $t$ points in the domain $\mathcal{L}$, say, $\mathcal{L}^*=\{ \bfell_1^*, \ldots, \bfell_t^*\}$, which is referred to as the \textit{reference set}. For instance, the reference set can be chosen to coincide with the set of input locations. Then we extend this finite-dimensional multivariate normal distribution to a stochastic process over the domain based on the reference set. 

To construct the NNGP, the process $w(\cdot)$ is assumed to depend on a few (say $t$) nearest neighbors in the reference set $\mathcal{L}^*$, where the nearness is defined according to some topological order for points in the reference set $\mathcal{L}^*$. Now define a \textit{history set} $\mathcal{H}(\bfell_i^*)$ of point $\bfell_i^*$, as follows: $\mathcal{H}(\bfell_1^*)$ is the empty set, and $\mathcal{H}(\bfell_i^*)\equiv \{ \bfell_1^*, \ldots, \bfell_{i-1}^*\}$ for $i=2, \ldots, t$. Then we define $N(\bfell_i^*)$ to be a neighbor set of $\bfell_i^*$, which is a subset of $\mathcal{H}(\bfell_i^*)$, and contains only a few points. Its definition is given as follows: 
\begin{equation} \label{eqn: neighbor sets}
N(\bfell_i^*) = \left \{  \begin{array}{ll}
\emptyset &\text{for }  i=1 \\
\mathcal{H}(\bfell_i^*) = \{ \bfell_1^*, \ldots, \bfell_{i-1}^*\} & \text{for } i=2, 3, \ldots, t \\
t \text{ nearest neighbors of } \bfell_i^* \text{ among } \mathcal{H}(\bfell_i^*) & \text{for } i=t+1, \ldots, n
\end{array}
\right.
\end{equation}

Based on the reference set equipped with the ordering mechanism specified above, the conditional distribution of $w(\bfell)$ is assumed to depend only on its neighbor sets $w(N(\bfell))$, where $N(\bfell)$ is a collection of $t$ nearest locations in the reference set $\mathcal{L}^*$. Given $N$ input locations $\mathcal{L}^O\equiv \{\bfell_1, \ldots, \bfell_n \}$, the joint distribution of $w(\cdot)$ over $\mathcal{L}^O$ is 
\begin{eqnarray*}
p(w(\mathcal{L}^O)) &=& p(w(\bfell_1)) \prod_{i=2}^n p(w(\bfell_i) \mid w(N(\bfell_i))) \\
&=& \mathcal{N}(w(\mathcal{L}^O) \mid \mathbf{0}, \widetilde{\mathbf{C}}),
\end{eqnarray*}
where $\widetilde{\mathbf{C}} :=(\mathbf{I} - \mathbf{A})^{-1}\mathbf{D} (\mathbf{I} - \mathbf{A})^{-\top}$. The matrix $\bfA$ is a strictly lower-triangular matrix with no more than $t$ elements in each row. The non-zero elements in the $i$th row are given by $\bfC_{N(\bfell_i), N(\bfell_i)}^{-1} \bfC_{\bfell_i, N(\bfell_i)}$. The matrix $\bfD$ is a diagonal matrix with $i$-th diagonal element given by $\sigma^2\{ C(\bfell_i, \bfell_i) - \bfC_{\bfell_i, N(\bfell_i)} \bfC_{N(\bfell_i), N(\bfell_i)}^{-1} \bfC_{N(\bfell_i), \bfell_i}\}$. Replacing the full covariance matrix $\mathbf{C}$ by $\widetilde{\bfC}$ in the likelihood function~\eqref{eqn: likelihood} leads to efficient computations. For detailed model formulation, see \cite{Datta2016}. In terms of posterior inference, \cite{Datta2016} use latent spatial variables to construct a sequential updating Markov chain Monte Carlo (MCMC) algorithm, which has  $O(nt^3)$ computational cost and $O(nt^2)$. As $t$ is much smaller than $n$, the computations in both model fitting and prediction are reduced tremendously. However, \cite{Finley2019} notice that the MCMC algorithm based on the spatial random variables has certain convergence issues (e.g., slow mixing), and \cite{Finley2019} propose the so-called collapsed MCMC algorithm which can deal with such issues. In Section~\ref{sec: Emulation} of the main text, we take a step further to integrate out both the spatial random variables and location-scale parameters in the NNGP model. Our proposed inference algorithm is very efficient, since only the correlation parameters and nugget parameters in the posterior distribution need to be updated with Metropolis-Hastings steps.  The nearest neighbor Gaussian process leads to a valid Gaussian process with a positive-definite covariance function. For notational convenience, we will refer to $\text{NNGP}(0, \sigma^2 c(\cdot, \cdot; \bftheta))$ as the NNGP process with the parent covariance function $\sigma^2 c(\cdot, \cdot; \bftheta)$. 

\section{PCA for Input Dimension Reduction} \label{app: PCA}
\begin{figure}[hbt!]
%\begin{center}
\makebox[\textwidth][c]{ \includegraphics[width=1.0\textwidth, height=0.2\textheight]{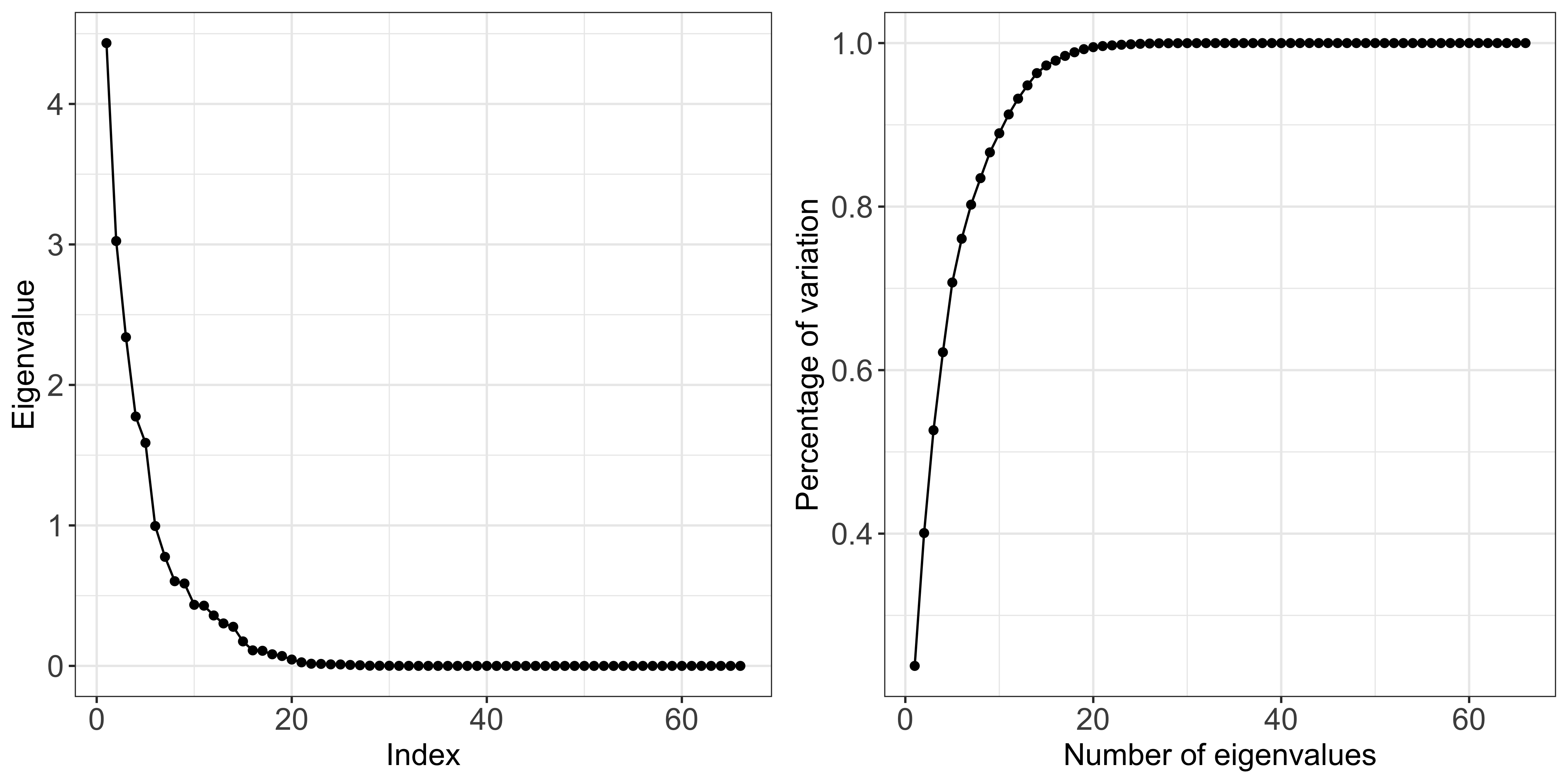}}
\caption{Diagnostics in the PCA method. The left panel shows the eigenvalues. The right panel shows the cumulative proportion of variation explained.}
\label{fig: PCA for inputs}
%\end{center}
\end{figure}

To reduce the dimensionality of the input space, we also use principal component analysis (PCA) on the standardized input data $\{ (\bfx_i, \bfb_i)\}_{i=1}^N$. Figure~\ref{fig: PCA for inputs} shows that the first 13 principal components can explain 94.8\% of the total variation. Notice that the PCA approach only uses information from input space for dimension reduction. In contrast, the active subspace method is a better alternative dimension reduction approach than the PCA approach in the OCO-2 application, since it incorporates gradient information of the FP forward model to perform dimension reduction. Our results suggest that only four active variables are needed to reduce the dimensionality of the space for state vectors, leading to 8 variables in Gaussian process emulation. This can save computational time substantially when fully Bayesian inference is employed. 

\section{Additional Numerical Illustrations} \label{app: numeric}
\begin{figure}[H]
    \captionsetup[figure]{justification=centering}
\makebox[\textwidth][c]{ \includegraphics[width=1.0\linewidth, height=0.2\textheight]{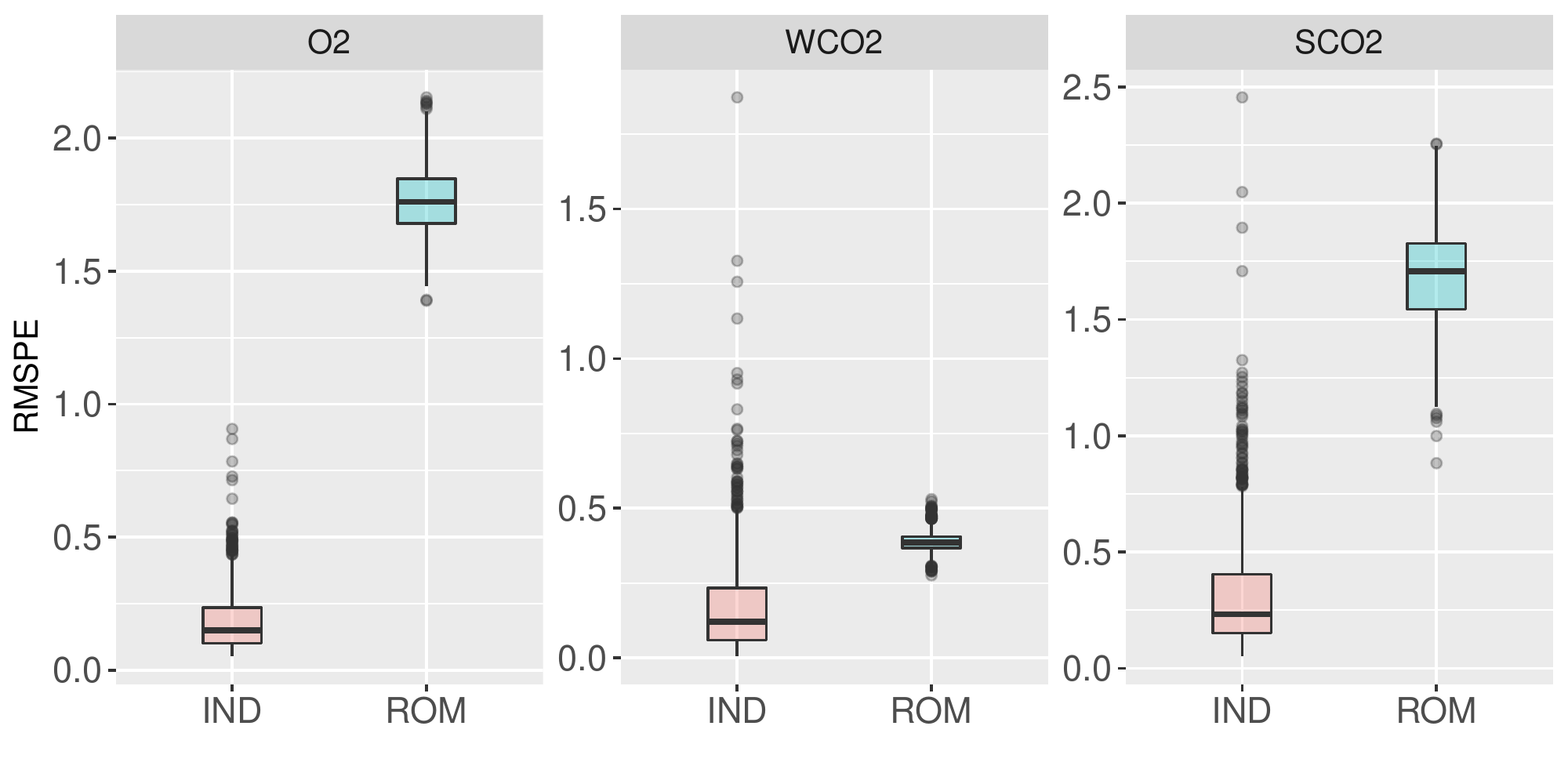}}
\caption{Box plot of pointwise RMSPE for predicted radiances based on the independent emulator (IND) and the reduced order model (ROM). The x-axis represents wavelength and the y-axis represents the RMSPE averaged over $n^*=1,000$ testing inputs and 1016 wavelengths.}
\label{fig: NNGP and ROM}
\end{figure}

\begin{figure}[H]
\makebox[\textwidth][c]{ \includegraphics[width=1.0\linewidth, height=0.35\textheight]{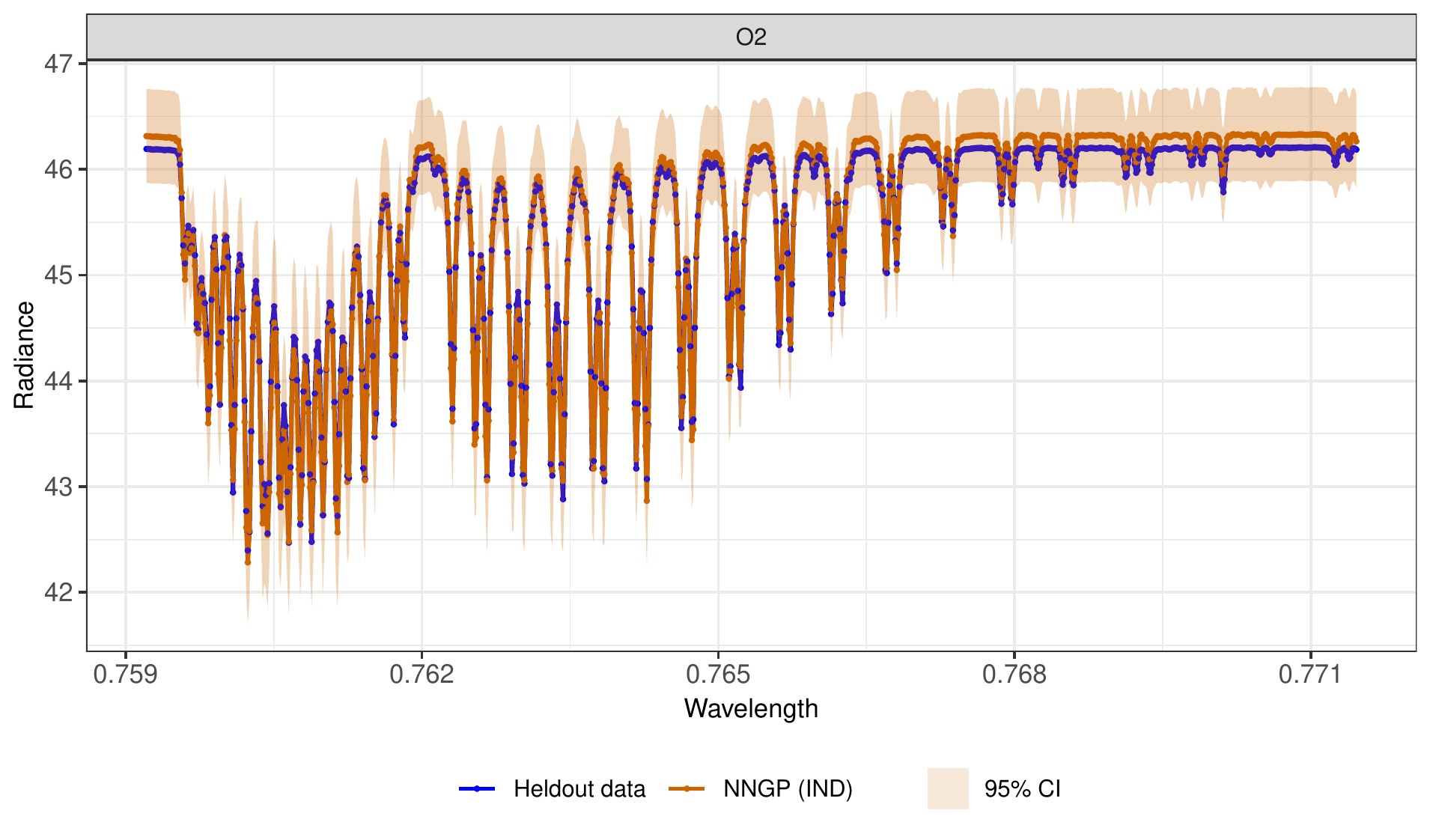}}
\caption{Comparison of the FP forward model simulated radiances and the predicted radiances with their 95\% credible intervals from the independent emulator (IND) for the O$_2$ band.}
\label{fig: O2 prediction}
\end{figure}

\begin{figure}[H]
\makebox[\textwidth][c]{ \includegraphics[width=1.0\linewidth, height=0.8\textheight]{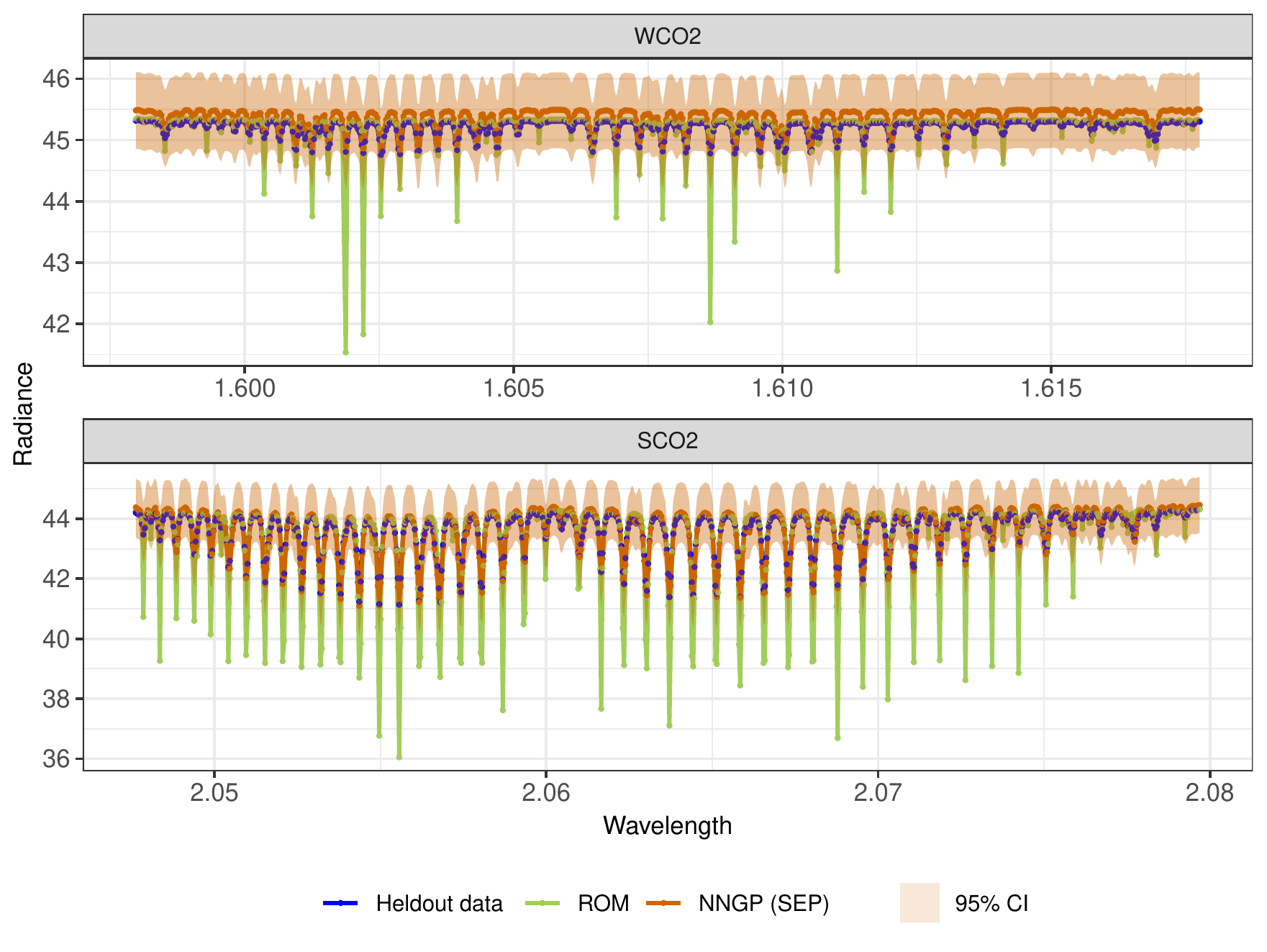}}
\caption{Comparison of the FP forward model simulated radiances, the ROM predictions, and the predicted radiances with their 95\% credible intervals from the separable emulator (SEP) for the WCO$_2$ and SCO$_2$ bands.}
\label{fig: SEP vs ROM}
\end{figure}

This section contains the addition numerical illustrations referenced in Section~\ref{sec: results}. Figure~\ref{fig: NNGP and ROM} shows box plots of RMSPE (averaged over all the wavelengths) across 1,000 testing inputs. We can see that the IND outperforms the ROM at each band in terms of RMSPE. For the O$_2$ band, the IND gives smaller RMSPE than the ROM at most of the testing inputs. For the WCO$_2$ and SCO$_2$ bands, the IND outperforms the ROM, with the exception that the former gives much larger RMSPE at a few soundings. Figure~\ref{fig: O2 prediction} shows that prediction for the O$_2$ band based on NNGP can capture the data much very well with good uncertainty quantification.  Figure~\ref{fig: SEP vs ROM} shows that the separable emulator can gives better prediction results than the ROM for the WCO$_2$ and SCO$_2$ bands with accurate uncertainty quantification.  

\begin{figure}[hbt!]
\begin{subfigure}{.33\textwidth}
  \centering
\makebox[\textwidth][c]{ \includegraphics[width=1.0\linewidth,height=0.2\textheight]{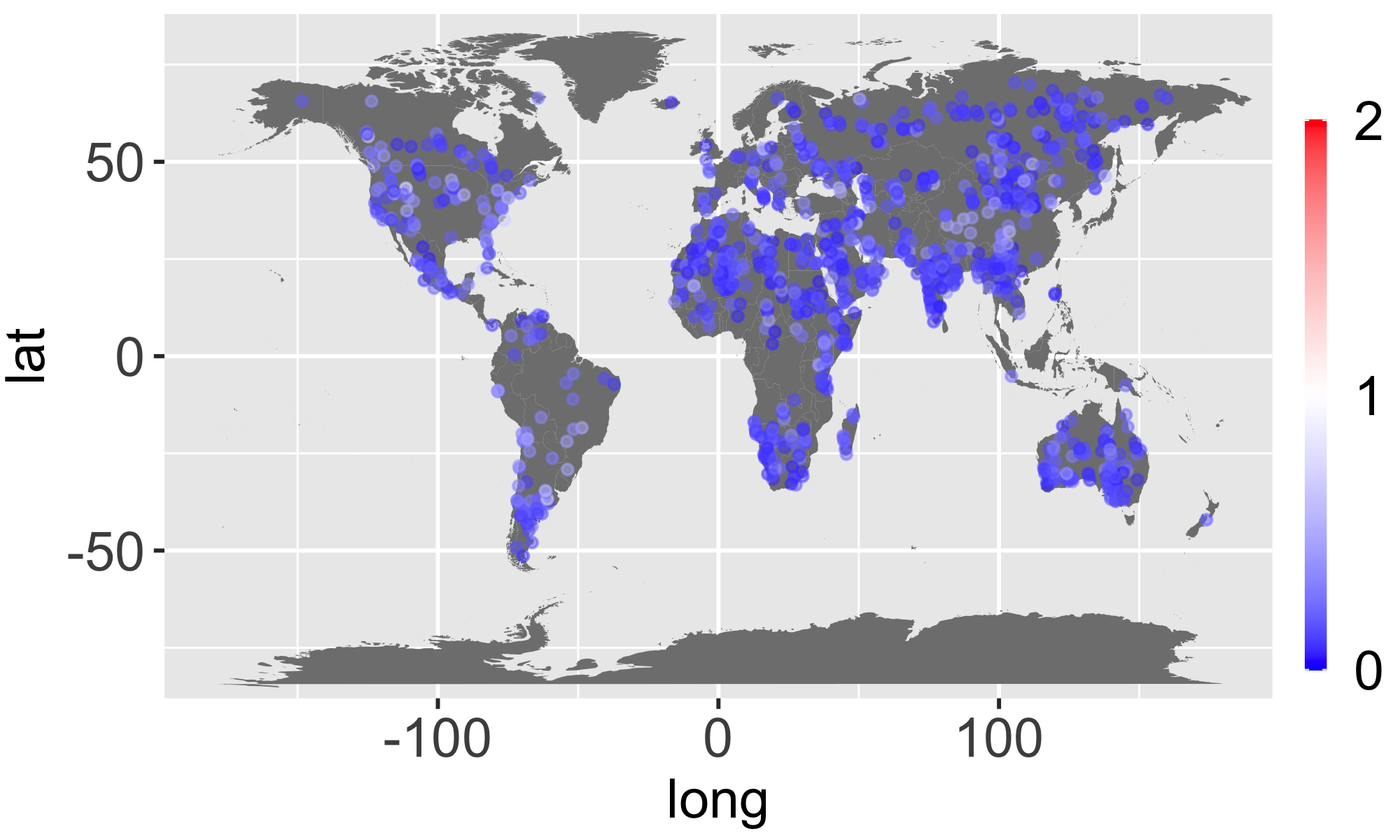}}
  \caption{IND: O$_2$ band}
  %\label{fig:}
\end{subfigure}%
\begin{subfigure}{.33\textwidth}
  %\centering
  \includegraphics[width=1.0\linewidth,height=0.2\textheight]{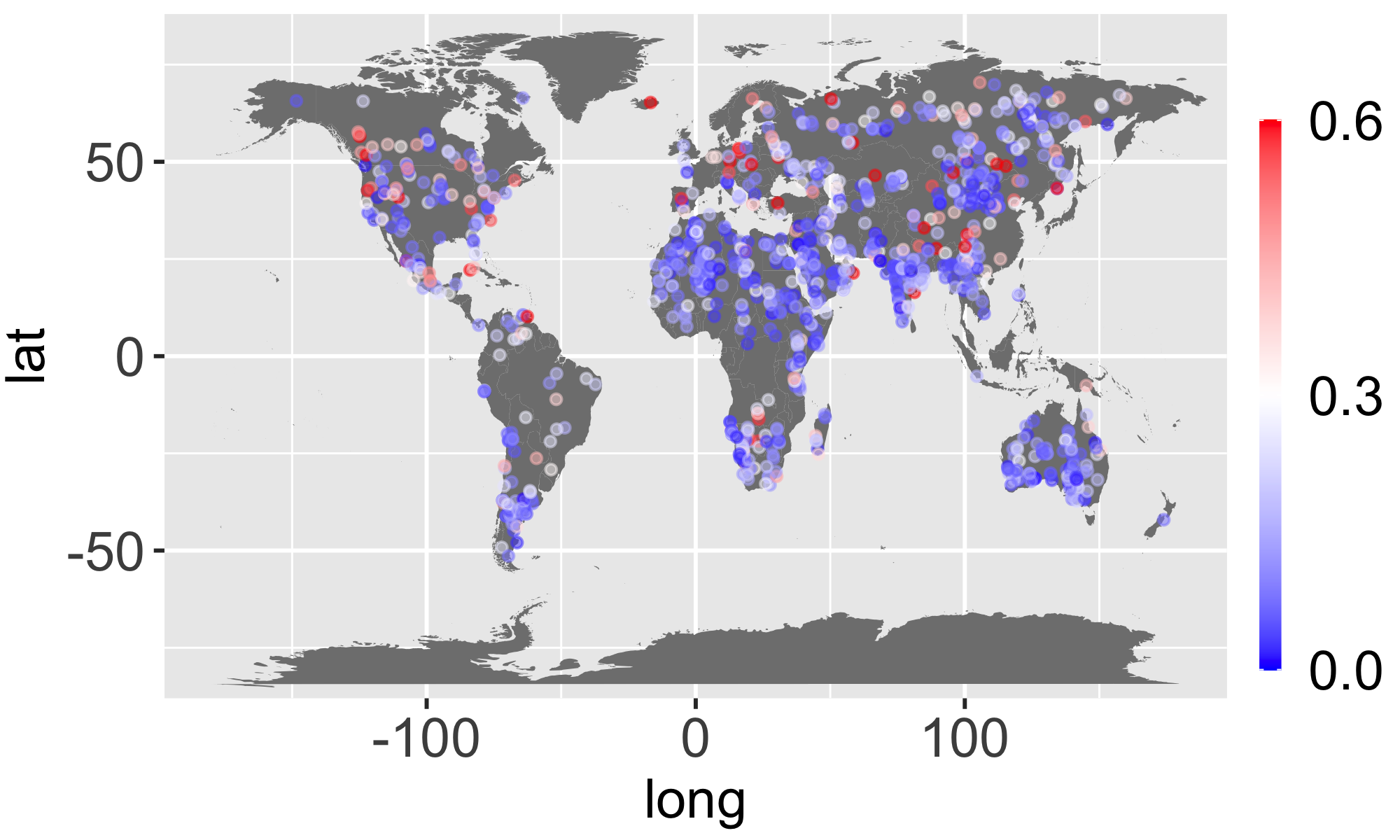}
  \caption{IND: WCO$_2$ band}
  %\label{fig:sfig2}
\end{subfigure}%
\begin{subfigure}{.33\textwidth}
  %\centering
  \includegraphics[width=1.0\linewidth,height=0.2\textheight]{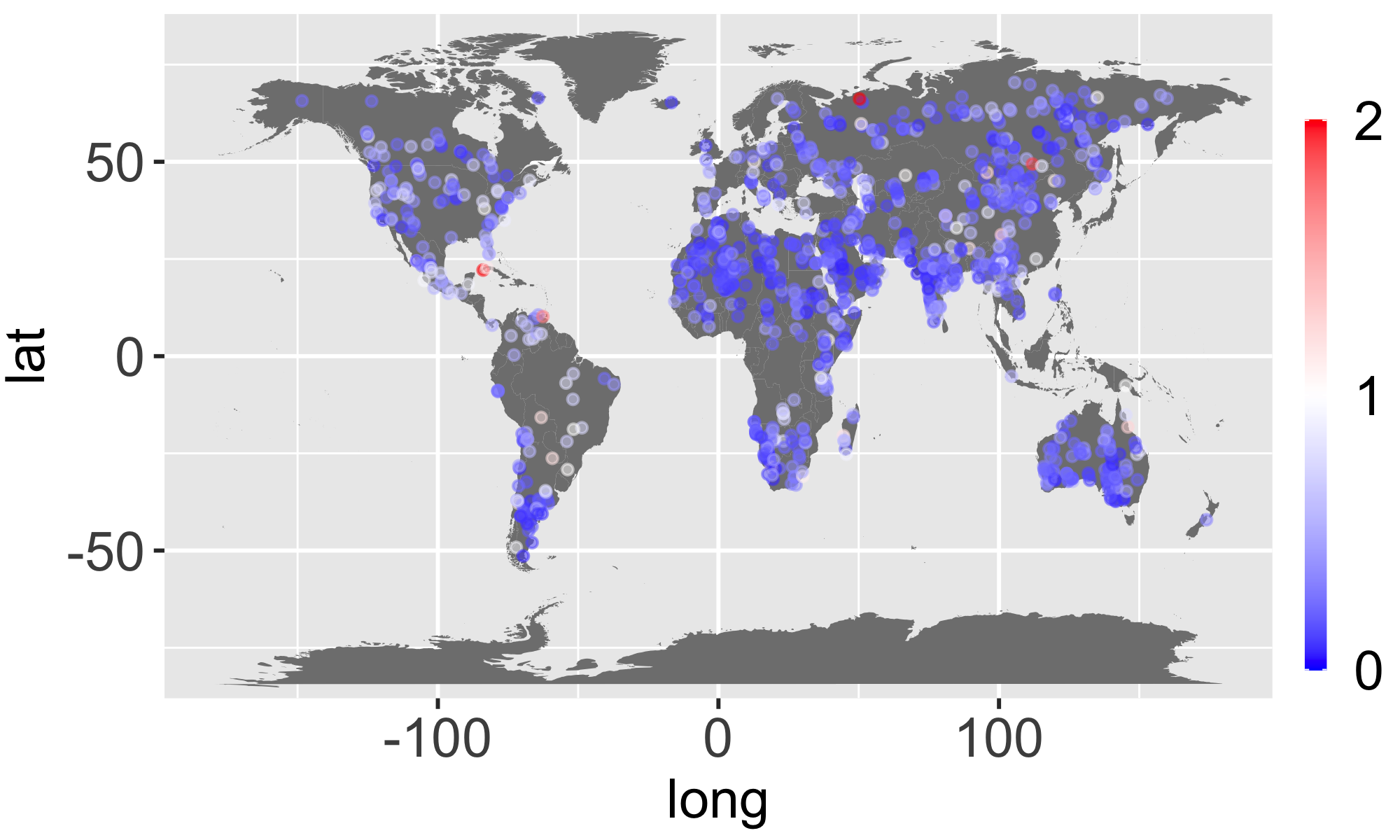}
  \caption{IND: SCO$_2$ band}
  %\label{fig:sfig2}
\end{subfigure}

\begin{subfigure}{.33\textwidth}
  \centering
\makebox[\textwidth][c]{ \includegraphics[width=1.0\linewidth,height=0.2\textheight]{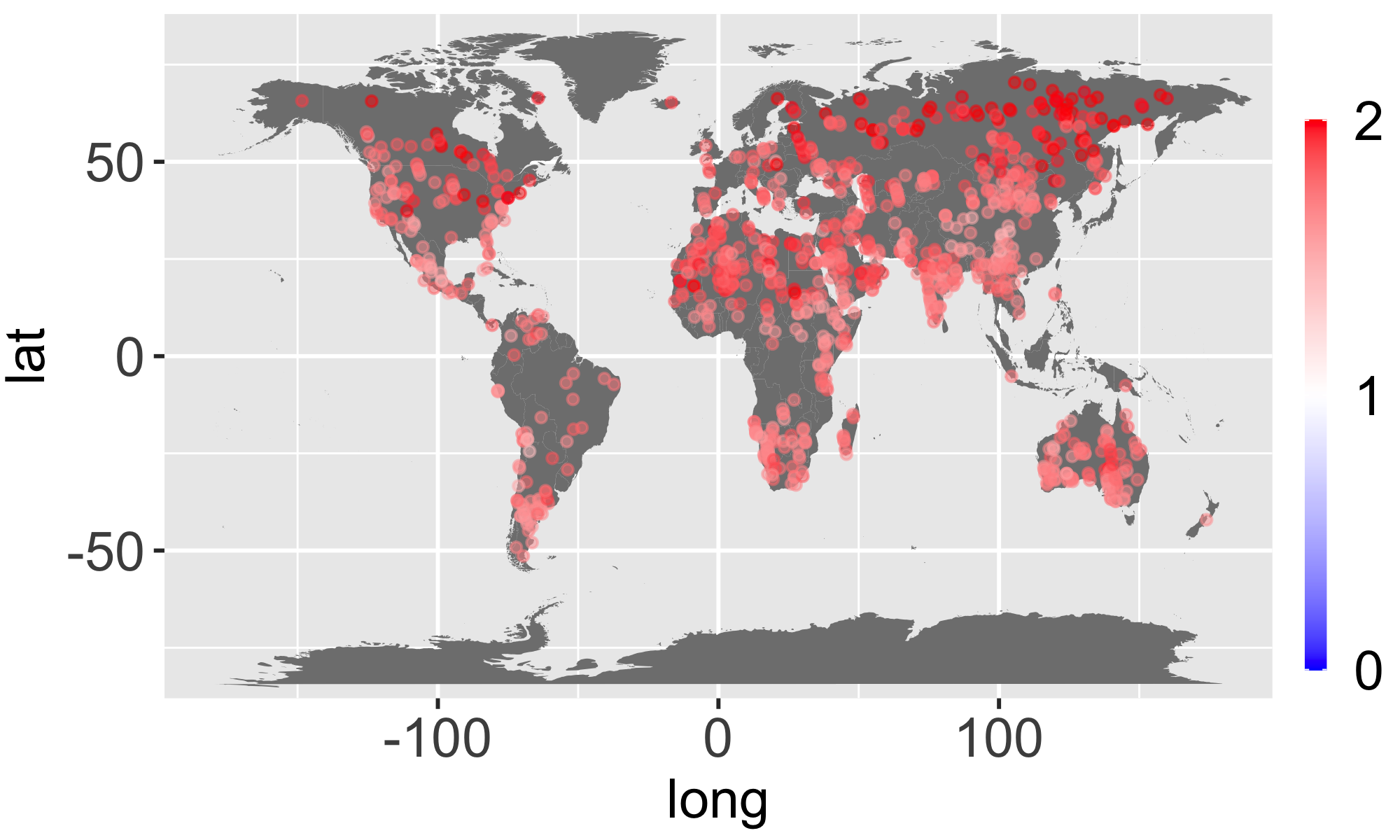}}
  %\caption{Difference between ROM and NNGP for O$_2$ band}
    \caption{ROM: O$_2$ band}
  %\label{fig:}
\end{subfigure}%
\begin{subfigure}{.33\textwidth}
  %\centering
  \includegraphics[width=1.0\linewidth,height=0.2\textheight]{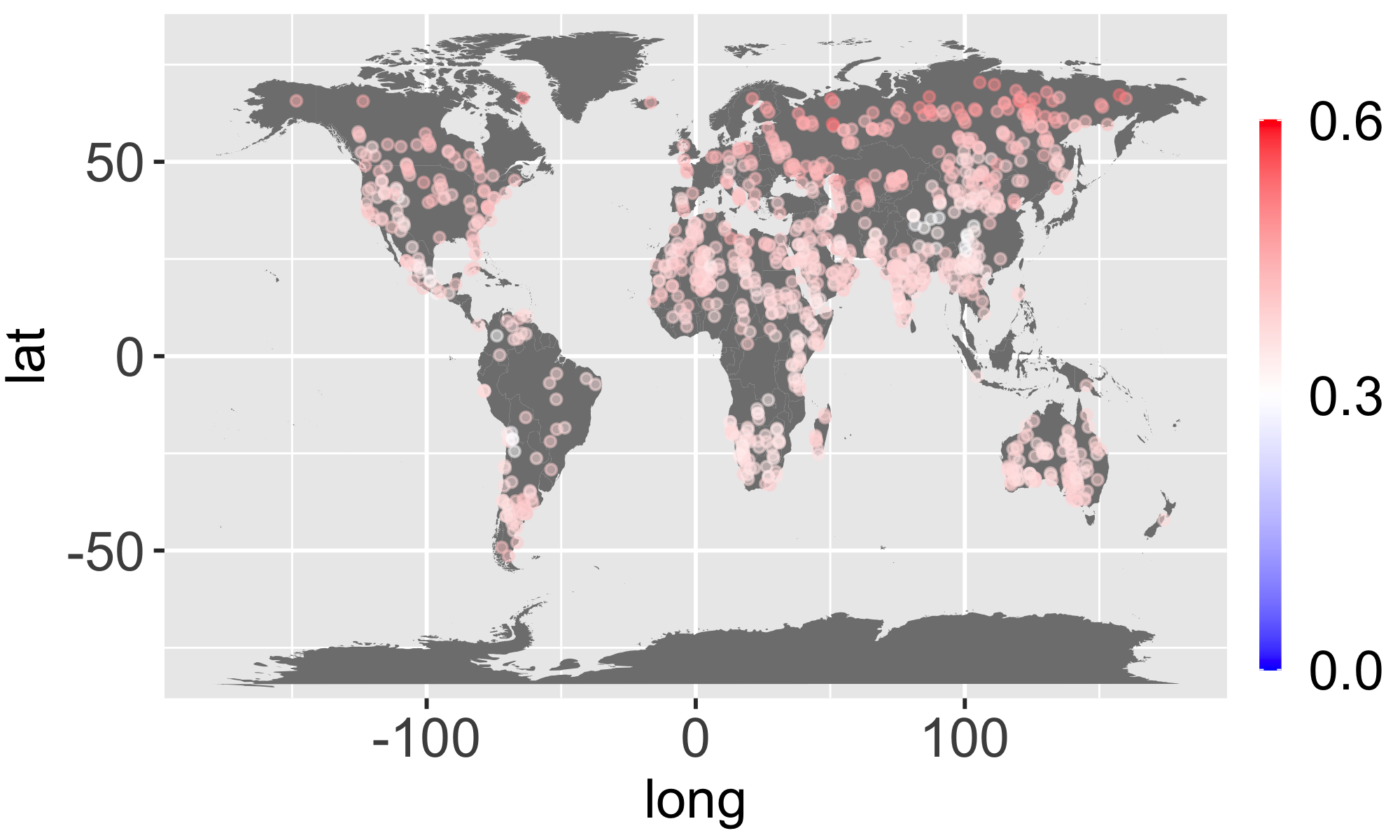}
  %\caption{Difference between ROM and NNGP for WCO$_2$ band}
    \caption{ROM: WCO$_2$ band}
  %\label{fig:sfig2}
\end{subfigure}%
\begin{subfigure}{.33\textwidth}
  %\centering
  \includegraphics[width=1.0\linewidth,height=0.2\textheight]{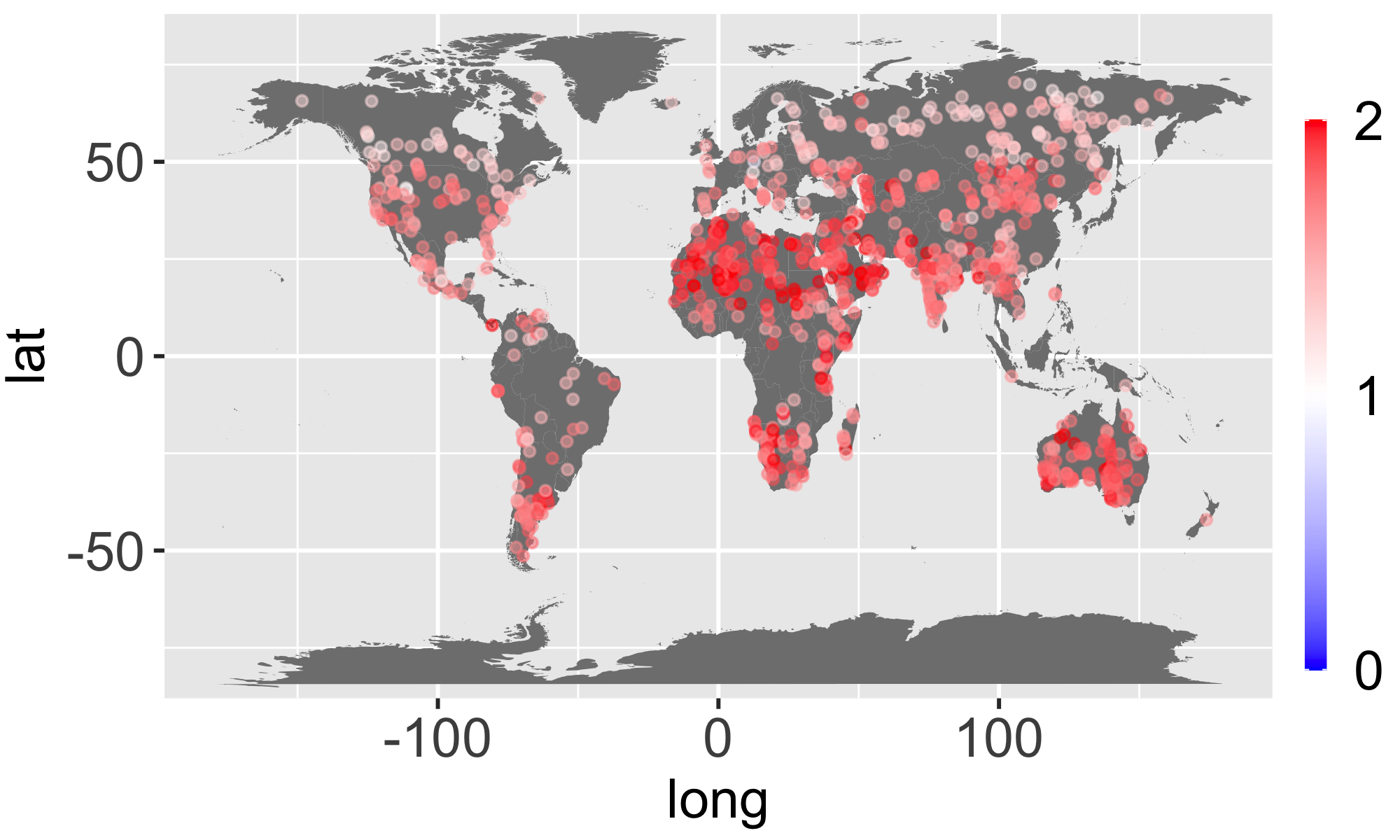}
  %\caption{Difference between ROM and NNGP for SCO$_2$ band}
    \caption{ROM: SCO$_2$ band}
  %\label{fig:sfig2}
\end{subfigure}

\caption{Spatial map of RMSPE based on the independent emulator (IND) and the reduced order model (ROM). The top panels show the averaged RMSPE over all wavelengths at each spatial location for the O$_2$ band (top-left panel), WCO$_2$ band (top-middle panel) and SCO$_2$ band (top-right panel) based on the IND. The bottom panels show the averaged RMSPE over all wavelengths at each spatial location for the O$_2$ band (top-left panel), WCO$_2$ band (top-middle panel) and SCO$_2$ band (top-right panel) based on the ROM.}
\label{fig: spatial map of NNGP and ROM}
\end{figure}

As each input is also associated with a longitude and latitude coordinate in space, we can also make visual comparison of the averaged RMSPE over all wavelengths between the IND and the ROM. Figure~\ref{fig: spatial map of NNGP and ROM} shows that the  O$_2$ band has comparatively smaller RMSPE than the WCO$_2$ and SCO$_2$ bands. There is no clear spatial pattern on large value of RMSPE across these three bands.  Besides, the ROM generally gives larger RMSPE than the emulator based on NNGP with exceptions at a few locations for the WCO$_2$ band. This agrees with previous results in Figure~\ref{fig: NNGP vs Surrogate}. The ROM seems to give larger RMSPE in the middle latitude bands than the IND. It is worth noting that the ROM can generate radiances for any given input, but there is no uncertainty associated with these radiances. Thus, this NNGP-based emulator also has an advantage over the ROM in terms of assessing prediction uncertainty.

%%%%%%%%%%%%%%%%%%%%%%%%%%%%%%%%%%
%%%%%%%%%%%%%%%%%%%%%%%%%%%%%%%%%%
%%%%% references

\end{document}